\documentclass[letterpaper,twocolumn,prl,aps,superscriptaddress,amsmath,amssymb,floatfix]{revtex4-1}
\usepackage{mathptmx}
\usepackage[latin9]{inputenc}
\usepackage{color}
\usepackage{amsmath}
\usepackage{amssymb}
\usepackage{graphicx}
\usepackage{esint}
\usepackage[unicode=true,
 bookmarks=true,bookmarksnumbered=false,bookmarksopen=false,
 breaklinks=false,pdfborder={0 0 1},backref=false,colorlinks=true]
 {hyperref}
\hypersetup{
 linkcolor=magenta,urlcolor=blue,citecolor=blue,pdfstartview={FitH},hyperfootnotes=false}

\makeatletter

\usepackage{textcomp}
\usepackage{epstopdf}

\pdfpageheight\paperheight
\pdfpagewidth\paperwidth

\@ifundefined{textcolor}{}{%
 \definecolor{BLACK}{gray}{0}
 \definecolor{WHITE}{gray}{1}
 \definecolor{RED}{rgb}{1,0,0}
 \definecolor{GREEN}{rgb}{0,1,0}
 \definecolor{BLUE}{rgb}{0,0,1}
 \definecolor{CYAN}{cmyk}{1,0,0,0}
 \definecolor{MAGENTA}{cmyk}{0,1,0,0}
 \definecolor{YELLOW}{cmyk}{0,0,1,0}
}

\usepackage{xcolor}\usepackage{soul}
\newcommand{\ket}[1]{\ensuremath{\left|#1\right\rangle}}

\definecolor{blue}{rgb}{0,0,1}
\definecolor{red}{rgb}{1,0,0}
\definecolor{green}{rgb}{0,1,0}

\usepackage{soul}

\makeatother
\begin{document}
	
\title{Dynamic compensation for pump-induced frequency shift in Kerr-cat qubit initialization}

\author{Yifang Xu}
\thanks{These authors contributed equally to this work.}
\author{Ziyue Hua}
\thanks{These authors contributed equally to this work.}
\author{Weiting Wang}
\email{wangwt2020@tsinghua.edu.cn}
\affiliation{Center for Quantum Information, Institute for Interdisciplinary, Information Sciences, Tsinghua University, Beijing 100084, China}

\author{Yuwei Ma}
\affiliation{Center for Quantum Information, Institute for Interdisciplinary, Information Sciences, Tsinghua University, Beijing 100084, China}

\author{Ming Li}
\affiliation{CAS Key Laboratory of Quantum Information, University of Science and Technology of China, Hefei 230026, China}

\author{Jiajun Chen}
\affiliation{Center for Quantum Information, Institute for Interdisciplinary, Information Sciences, Tsinghua University, Beijing 100084, China}

\author{Jie~Zhou}
\affiliation{Center for Quantum Information, Institute for Interdisciplinary, Information Sciences, Tsinghua University, Beijing 100084, China}

\author{Xiaoxuan~Pan}
\affiliation{Center for Quantum Information, Institute for Interdisciplinary, Information Sciences, Tsinghua University, Beijing 100084, China}

\author{Lintao~Xiao}
\affiliation{Center for Quantum Information, Institute for Interdisciplinary, Information Sciences, Tsinghua University, Beijing 100084, China}

\author{Hongwei~Huang}
\affiliation{Center for Quantum Information, Institute for Interdisciplinary, Information Sciences, Tsinghua University, Beijing 100084, China}

\author{Weizhou Cai}
\affiliation{CAS Key Laboratory of Quantum Information, University of Science and Technology of China, Hefei 230026, China}

\author{Hao Ai}
\affiliation{Institute of Microelectronics, Tsinghua University, Beijing 100084, China}

\author{Yu-xi Liu}
\affiliation{Institute of Microelectronics, Tsinghua University, Beijing 100084, China}

\author{Chang-Ling Zou}
\email{clzou321@ustc.edu.cn}
\affiliation{CAS Key Laboratory of Quantum Information, University of Science and Technology of China, Hefei 230026, China}
\affiliation{Hefei National Laboratory, Hefei 230088, China}

\author{Luyan Sun}
\email{luyansun@tsinghua.edu.cn}
\affiliation{Center for Quantum Information, Institute for Interdisciplinary, Information Sciences, Tsinghua University, Beijing 100084, China}
\affiliation{Hefei National Laboratory, Hefei 230088, China}


\begin{abstract}
The noise-biased Kerr-cat qubit is an attractive candidate for fault-tolerant quantum computation; however, its initialization faces challenges due to the squeezing pump-induced frequency shift (PIFS). Here, we propose and demonstrate a dynamic compensation method to mitigate the effect of PIFS during the Kerr-cat qubit initialization. Utilizing a novel nonlinearity-engineered multi-loop SQUID device, we realize a stabilized Kerr-cat qubit and validate the advantages of the dynamic compensation method by improving the initialization fidelity from 57\% to 78\%, with a projected fidelity of 91\% after excluding state preparation and measurement errors. Our results not only advance the practical implementation of Kerr-cat qubits, but also provide valuable insights into the fundamental adiabatic dynamics of these systems. This work paves the way for scalable quantum processors that leverage the bias-preserving properties of Kerr-cat qubits.
\end{abstract}
\maketitle

\textit{Introduction.-} Bias-preserving cat qubits~\cite{DCMirrahimi2014NJP,DCZaki2015Science}, which encode quantum information in Schr\"odinger cat states, have emerged as a promising platform for fault-tolerant quantum computation~\cite{BPAliferis2008PRAFaultTolerant,BPAliferis2009NJPFaultTolerantCase,BPWebster2015PRA,BPChamberland2022PRXQFaultToleDissCat}. These qubits offer exponential suppression of bit-flip errors~\cite{DCMirrahimi2014NJP,DCLescanne2020NP,KCPuri2017npjQI} and reduced resource overhead for quantum error correction. The bias-preserving property of cat qubits is particularly advantageous for implementing surface codes~\cite{SurfaceCodeFowler2012PRL,XZZXCodeAtaides2021NC,BPTuckett2018PRLBiasQubitSurface,BPTuckett2019PRXBiasQubitSurface,BPTuckett2020PRLBiasQubitSurface,BPDarmawan2021PRXQXZZXKerrCat}, repetition codes~\cite{Shor1995PRAcode,Steane1996PRLcode,DCGuillaud2019PRXRepetition,DCGuillaud2021PRARepetitionThreshold,DCGouzien2023PRLRepetitiond5}, and error syndrome detection~\cite{KCPuri2019PRXDetector,KCAndy2024arxivDetector}. Exciting experimental progress has been achieved recently, with cat qubit lifetimes exceeding 100 seconds without quantum control~\cite{DCBerdou2023PRXQ100s} and 10 seconds with quantum control~\cite{DCReglade2024Nature10sControl}.  

Two main approaches exist for realizing cat qubits: dissipative cat qubits~\cite{DCMirrahimi2014NJP,DCLescanne2020NP} and Kerr-cat qubits~\cite{KCPuri2017npjQI,KCGrimm2020Nature}. Kerr-cat qubits are particularly appealing due to their potential for high-speed gate operations~\cite{KCXu2022PRRKCatFast}. However, their initialization, which typically involves adiabatically ramping up a two-photon pump to map the Fock states to the cat state manifold~\cite{KCPuri2017npjQI,KCGrimm2020Nature}, faces a significant challenge: the pump-induced frequency shift (PIFS). This shift, arising from either charge~\cite{KCGrimm2020Nature} or flux~\cite{FPLu2023NCBeamSplitting,KCIyama2024NCWigner,KCHoshiArxiv2KPO} pumping, can cause decoherence and state leakage during the initialization~\cite{KCGrimm2020Nature,KCPuri2020SciAdvBPCX}. Previous attempts to compensate for PIFS rely on applying a static detuning between the pump and qubit frequencies~\cite{KCGrimm2020Nature}. While this approach works for small PIFS, it fails to account for the dynamic nature of PIFS during the ramping-up process, leading to imperfect initialization. Moreover, as the pursuit of better bias-preserving properties necessitates larger cat amplitudes, the effect of PIFS becomes increasingly pronounced and demands a more sophisticated compensation strategy.

In this work, we introduce a dynamic compensation method that continuously adjusts the pump frequency to match the instantaneous PIFS during the adiabatic initialization. 
We experimentally demonstrate the advantage of this dynamic compensation for a Kerr-cat qubit using a novel Nonlinearity-Engineered Multi-loop SQUID (NEMS) device~\cite{NEMS}. This device not only holds promise for realizing bias-preserving CNOT gates~\cite{KCPuri2020SciAdvBPCX,BPChamberland2022PRXQFaultToleDissCat}, but also offers independent tuning of nonlinearities compared with other Kerr-cat qubit candidates such as SNAIL~\cite{SNAILFrattini2017APL3waveMixing,SNAIL2018FrattiniPRATUPA,SNAILSivak2019PRAKerrFree,KCHe2023NCK0}, ATS~\cite{DCLescanne2020NP}, Josephson ring modulator~\cite{JRMBergeal2010NP,JRMRoch2012PRL}, and a series of asymmetric DC SQUIDs~\cite{KCIyama2024NCWigner,KCHoshiArxiv2KPO}. Our theoretical and experimental results showcase the superiority of the  dynamic compensation approach, particularly for large PIFS values. Our demonstration paves the way for high-fidelity Kerr-cat qubit operations and advances the field towards fault-tolerant quantum computation. 

\begin{figure}[t]
	\centering
	\includegraphics{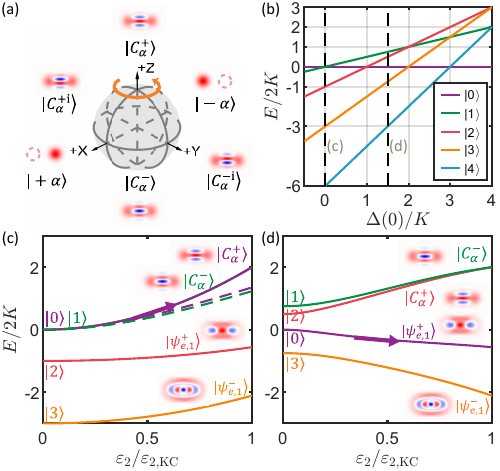}
	\caption{(a) Bloch sphere representation of the Kerr-cat qubit encoding space. Here, $\ket{\pm\alpha}$ denotes the coherent states and $\ket{C_\alpha^k} = (\ket{\alpha} + k\ket{-\alpha})/\sqrt{2}$ for $k=+,-,+\mathrm{i},-\mathrm{i}$. (b) Energy levels of the system as a function of $\Delta(0)$, the detuning at $t=0$ in Eq.~\ref{eq:HCat} with no external pump. (c) Solid lines: evolution of the energy levels for an ideal case with $\Delta\equiv0$ when ramping up the pump amplitude $\epsilon_2$ to a final value of $\epsilon_\mathrm{2,KC}$. Dashed lines: the removal of the degeneracy due to the pump-induced frequency shift (PIFS). (d) Evolution of the energy levels for a static compensation with a fixed pump detuning $\Delta_\mathrm{as}(t)\equiv 1.5K$. At $\epsilon_2=0$, $\Delta(0)=1.5 K$, corresponding to the right dashed line in (b). 
    In (c) and (d), the purple arrows indicate adiabatic evolution paths that map the vacuum state to $\ket{C_\alpha^+}$ and to $\ket{\psi_\mathrm{e,1}^+}\approx[\hat{D}(\alpha)+\hat{D}(-\alpha)]\ket{1}$, respectively.
    } 
	\label{fig:1}  
\end{figure}

\textit{Principle.-} The Kerr-cat qubit can be realized with a single-bosonic mode Hamiltonian ($\hbar=1$)~\cite{KCPuri2017npjQI,KCGrimm2020Nature}
\begin{equation}\label{eq:HCat}
	\hat{H}_\mathrm{cat} = \Delta(t) a^{\dagger}a-K\hat{a}^\dagger\hat{a}^\dagger\hat{a}\hat{a} + 
 \varepsilon_2(t)(\hat{a}^{\dagger2}+\hat{a}^2),
\end{equation}
where $\hat{a}^\dagger(\hat{a})$ is the creation (annihilation) operator of the mode, $\Delta$ is the relative frequency detuning, $K$ is the Kerr nonlinearity, and $\varepsilon_2$ is real and represents the pump amplitude. In an ideal experimental configuration with $\Delta=0$, the bosonic mode has a stabilized two-dimensional Hilbert space spanned by $\ket{\pm\alpha}$, i.e., the coherent states with an average photon number $|\alpha|^2=\varepsilon_2/K$. Consequently, the superposition states $\ket{C_\alpha^k} = (\ket{\alpha} + k\ket{-\alpha})/\sqrt{2}$ for $k=+,-,+\mathrm{i},-\mathrm{i}$  are also stabilized states in the mode. Figure~\ref{fig:1}(a) presents the Bloch sphere of the Kerr-cat qubit by treating $\ket{C_\alpha^\pm}$ as the computational basis. The Kerr-cat qubit has the advantage of being bias-preserving, with the states along the X-axis having an enhanced lifetime that increases exponentially with the Kerr-cat size $|\alpha|^2$~\cite{KCPuri2017npjQI,KCPuri2020SciAdvBPCX}. 

\begin{figure*}[t]
	\centering
	\includegraphics{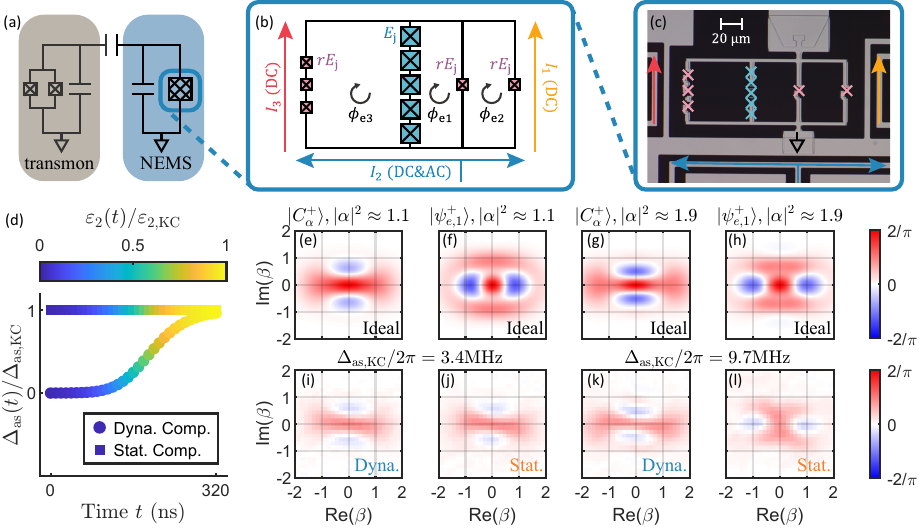}
	\caption{
	(a) Circuit schematic of the device. The Nonlinearity-Engineered Multi-loop SQUID (NEMS) is shunted with a large capacitor, similar to that of a transmon.
	(b) Circuit schematic of the NEMS. $I_2$ is the flux pumping port. 
    The loop areas are meticulously designed such that the flux ratios are 1:-1:3, and
    the ratio of the small (pink) to the large (blue) Josephson junctions is $r\approx0.16$. More details and parameters about the device are provided in Ref.~\cite{SM}. 
	(c) Photograph of the NEMS. 
	(d) Illustration comparing the dynamic and static compensation methods. The envelope $\varepsilon_2(t)$ follows a tanh function. In the static method, the pump frequency remains fixed over time, whereas in the dynamic method, the pump frequency varies with time $\Delta_\mathrm{as}(t)\propto\varepsilon^2_2(t)$. 
	(e-l) Comparison of the ideal and experimental Wigner functions for two typical PIFS.
	(e-h) The first two even-parity eigenstates for $|\alpha|^2\approx1.1$ and 1.9, with $\Delta_{\mathrm{as,KC}}=3.4\,\mathrm{MHz}<K$ and $\Delta_{\mathrm{as,KC}}=9.7\,\mathrm{MHz}>K$.
	(i-l) The experimental results based on the dynamic and static methods. 
	The color contrast is mainly limited by the decoherence of the transmon.
	}
	\label{fig:2}  
\end{figure*}

The initialization or encoding of the Kerr-cat qubit can be realized by ramping up the amplitude of the squeezing pump $\varepsilon_2$. As shown by the solid lines in Fig.~\ref{fig:1}(c), the eigenstates of the Hamiltonian evolve from the Fock state basis \{$\ket{0}$, $\ket{1}$\} to the computational basis of the Kerr-cat qubit following the traces as $\varepsilon_2$ increases. However, in practical experiments there are parasitic effects on the mode when applying pumps to it. Specifically, we encounter a dynamic PIFS $\Delta_\mathrm{PI}(t)=\gamma\varepsilon_2^2(t)$ with a coefficient $\gamma$. Consequently, the corresponding detuning varies with the pump as $\Delta=\Delta_{\mathrm{as}} -\Delta_\mathrm{PI}$, where $\Delta_\mathrm{as} = \omega_\mathrm{a} - \omega_\mathrm{s}/2$ is the detuning between the external squeezing pump ($\omega_{\mathrm{s}}$) and the bare mode ($\omega_{\mathrm{a}}$). 
We define the pump amplitude at the end of the ramping-up time $t_\mathrm{up}$ as $\varepsilon_2(t_\mathrm{up})=\varepsilon_\mathrm{2,KC}$ and the detuning $\Delta_\mathrm{as}(t_\mathrm{up})=\Delta_\mathrm{as,KC}$. Then, $\Delta_\mathrm{PI}(0)=0$ and $\Delta_\mathrm{PI}(t_\mathrm{up})=\Delta_\mathrm{PI,KC}=\gamma\varepsilon^2_\mathrm{2,KC}$. Here the subscript KC denotes the parameters for preserving the Kerr-cat qubit.

If the PIFS is not compensated, i.e., $\Delta_{\mathrm{as}}\equiv0$, then $\Delta=0$ holds true only at $t=0$.
As $\Delta$ increases when ramping up the pump, the corresponding eigenenergies of the Hamiltonian become non-degenerate, as shown by the dashed lines in Fig.~\ref{fig:1}(c) with $\Delta_{\mathrm{PI,KC}}=0.8K$ as an example. Additionally, the prepared Kerr-cat qubit at $t>t_{\mathrm{up}}$ experiences a non-zero $\Delta$, which induces a Z-rotation of the Kerr-cat qubit [orange arrow in Fig.~\ref{fig:1}(a)] with a precession rate $\Omega_Z = 4\Delta|\alpha|^2\exp(-2|\alpha|^2)$ ~\cite{KCGrimm2020Nature,KCPuri2020SciAdvBPCX}.  This rotation causes the state in the well-protected axis to pass through the not-protected axis, violating the principle of bias-preserving qubits and significantly shortening the coherence time.

A simple solution to this problem is to introduce a static detuning of the pump to compensate $\Delta_{\mathrm{PI,KC}}$ after the initialization, i.e., $\Delta_{\mathrm{as}}(t)\equiv\Delta_{\mathrm{as,KC}}=\Delta_{\mathrm{PI,KC}}$.
However, this static compensation lifts the degeneracy of Fock states at the beginning of the initialization due to $\Delta(0)=\Delta_\mathrm{as,KC}$. As illustrated in Fig.~\ref{fig:1}(b), the eigenenergies of the eigenstates (Fock states) at $t=0$ vary with $\Delta(0)$, resulting in a rearrangement of the order of the Fock states. Similar results are also discussed in Ref.~\cite{StatePreYanagimoto2019PRAFock}. When ramping up $\varepsilon_2$, as shown in Fig.~\ref{fig:1}(d) for $\Delta_{\mathrm{PI,KC}}=1.5K$ as an example, the evolution of eigenstates is significantly different from that in Fig.~\ref{fig:1}(c). 

Two potential issues arise for such a static compensation. First, the traces of the eigenstates change significantly, and we should choose higher photon number states for adiabatic initialization of the target cat states. Otherwise, we end up with a leakage state [purple arrow in Fig.~\ref{fig:1}(d)]. 
Second, the energy gaps between different traces are reduced as compared to the ideal cases in Fig.~\ref{fig:1}(c), imposing stringent adiabatic ramping rate, i.e., $t_\mathrm{up}\gg1/\delta_E$, where $\delta_E$ is the smallest energy gap~\cite{Dirac1925PRSAdiabatic}. 
Considering the relationship $\Delta_{\mathrm{PI,KC}}=\gamma |\alpha|^4 K^2$, such effects cannot be neglected when attempting to achieve excellent bias noise properties by increasing $|\alpha|^2$. 

Therefore, we propose an alternative approach that dynamically varies the pump frequency to compensate the PIFS while maintaining $\Delta=0$ throughout the entire ramping process. This approach effectively recovers the ideal eigenenergy traces shown in Fig.~\ref{fig:1}(c), enabling high-fidelity initialization of the Kerr-cat qubit. 

\textit{Experiments.-} The Kerr-cat qubit is implemented using a novel NEMS device with its schematic circuit shown in Fig.~\ref{fig:2}(a). An ancillary transmon qubit is introduced for characterizing the Kerr-cat qubit. The NEMS consists of three loops [Fig.~\ref{fig:2}(b)], holding the advantage of independently tuning the nonlinearities of different orders by controlling the external flux through three current lines $I_1$, $I_2$, and $I_3$~\cite{NEMS,SM}. Figure~\ref{fig:2}(c) depicts the fabricated NEMS device, with the ratios of the junctions and loop areas being meticulously designed. By biasing the NEMS at $(\phi_\mathrm{e1},\phi_\mathrm{e2},\phi_\mathrm{e3}) = (-0.08,1.08,0)\pi$ and applying a flux pump via $I_2$ line, the Hamiltonian of Eq.~\ref{eq:HCat} for the Kerr-cat qubit can be realized~\cite{SM}. 

Without an external pump, the NEMS functions as a nonlinear oscillator at a frequency of $\omega_\mathrm{a}/2\pi =5.6\,\mathrm{GHz}$ with a Kerr nonlinearity $K/2\pi = 6.9\,\mathrm{MHz}$. The amplitude damping and transverse coherence times are $T_\mathrm{1N} = 6.0\,\mathrm{\mu s}$ and $T_\mathrm{2N} = 3.0\,\mathrm{\mu s}$, respectively. 
The Kerr-cat qubit is realized by applying a flux pump at $\omega_\mathrm{s}/2\pi\approx11.2\,\mathrm{GHz}$ with an amplitude of $\varepsilon_\mathrm{2,KC}/2\pi=9.7\,\mathrm{MHz}$ to the NEMS. 
By initializing the NEMS in the state $(\ket{0}+\ket{1})/\sqrt{2}$, the Kerr-cat qubit can be prepared in the $\ket{\alpha}$ state on the equator of the Bloch sphere. The precession due to PIFS is measured by varying the holding time of NEMS in the Kerr-cat states, allowing the calibration of $\Delta_{\mathrm{as,KC}}$~\cite{SM}. Furthermore, the Rabi drive of the Kerr-cat qubit is realized by a charge drive on the NEMS, with a Rabi frequency along the X-axis given by $\Omega_X = \mathrm{Re}(4\epsilon_\mathrm{x}\alpha)/\sqrt{1-\exp(-4|\alpha|^2)}$ ~\cite{KCPuri2020SciAdvBPCX}. Here, $\epsilon_\mathrm{x}$ is the Rabi drive strength. By fitting the Rabi dynamics, the Kerr-cat size is calibrated as $|\alpha|^2=1.41$, with a significantly extended damping time of $T_\mathrm{1,KC} = 13\,\mathrm{\mu s}$~\cite{SM}.

Next, we systematically investigate the performance of the static and dynamic compensation methods in mapping between the Fock states and Kerr-cat qubit states. 
Specifically, we focus on the mapping from $\ket{0}$ to $\ket{C_\alpha^+}$, since this mapping is not subject to gate fidelity in the Fock space.
Figure~\ref{fig:2}(d) illustrates the frequency (vertical axis) and amplitude (color) profiles for the ramping-up process, comparing the two methods. Both schemes utilize a tanh function for the amplitude $\varepsilon_2$, while the dynamic compensation method employs a tanh-squared function for the frequency, with a ramping-up time $t_\mathrm{up}=320~\mathrm{ns}$ that satisfies the adiabatic condition $t_\mathrm{up}\gg1/2K$~\cite{KCGrimm2020Nature}.
The experimental Wigner functions~\cite{ParitySun2014Nature,KCIyama2024NCWigner} of the Kerr-cat state with $|\alpha|^2\approx1.1$ and 1.9, using different compensation schemes, are shown in Figs.~\ref{fig:2}(i)-(l).
For a small $|\alpha|^2$, where the PIFS is smaller than $K$, both methods yield the same results [Figs.~\ref{fig:2}(i) and \ref{fig:2}(j)] that match the target state $\ket{C_\alpha^+}$ [Fig.~\ref{fig:2}(e)] but are distinct from the potential leakage state $\ket{\psi_{e,1}^+}$ [Fig.~\ref{fig:2}(f)]. However, when the PIFS is larger than $K$ but smaller than $3K$, i.e., $1.4K$ for $|\alpha|^2=1.9$, the dynamic approach outperforms the static approach. As shown in Fig.~\ref{fig:2}(k), the dynamic compensation method yields a state that more closely resembles the ideal eigenstate [Fig.~\ref{fig:2}(g)]. In contrast, the static compensation method leads the state [Fig.~\ref{fig:2}(l)] to deviate significantly from the ideal case and appear more similar to the leakage state $\ket{\psi_\mathrm{e,1}^+}$ [Fig.~\ref{fig:2}(h)].
This confirms that the static compensation fails to account for the time-dependent nature of the PIFS, leading to imperfect initialization and increased leakage as the PIFS grows larger. 

\begin{figure}[t]
	\centering
	\includegraphics{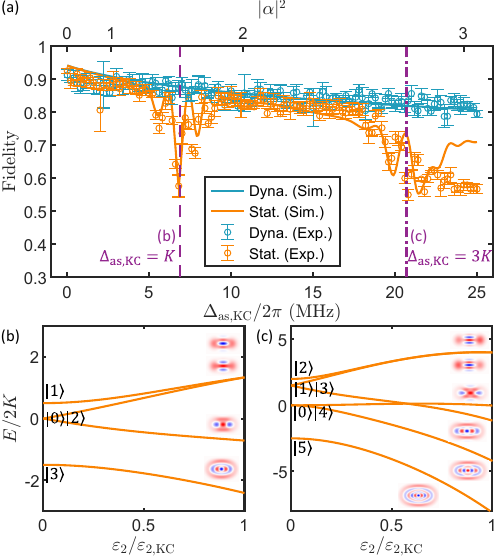}
	\caption{(a) Experimental process fidelities of the combined ramping-up and ramping-down operations for different pump strengths, corresponding to various compensation frequencies ($\Delta_{\mathrm{as,KC}}$) and Kerr-cat sizes ($|\alpha|^2$), using the dynamic and static compensation methods. Vertical dashed lines: the coalescence of Fock state energies around $\Delta_{\mathrm{as,KC}}=K$ and $3K$. The data points are averaged over five repeated measurements, with error bars indicating the standard deviation. (b) and (c) Evolution of the energy levels for the static compensation method with $\Delta_{\mathrm{as,KC}}=K$ and $3K$, respectively, illustrating the Landau-Zener transition for Fock states $\{\ket{0},\ket{2}\}$ and $\{\ket{1},\ket{3}\}$.  
	 }
	\label{fig:3}  
\end{figure}

The fidelities of the mapping based on the two schemes are further characterized by implementing a symmetric ramping-down process to remap the Kerr-cat states back to the Fock space. This whole process is ideally an identity channel, and can be directly characterized by performing quantum process tomography (QPT)~\cite{NielsenChuang} in the Fock state basis when the pump is off. Figure~\ref{fig:3}(a) presents the process fidelity for different pump amplitudes, i.e., different PIFS and $|\alpha|^2$, obtained using the two schemes. It demonstrates that the dynamic compensation method consistently outperforms the static compensation method across the entire range of PIFS values. In particular, when $\Delta_{\mathrm{as,KC}}$ is near $K$ or $3K$, the static compensation method shows strong oscillations of the fidelity. This can be explained by the collapse of the energy levels of the Fock states with respect to the compensated detunings, and the condition for the adiabatic evolution is violated. 
This leads to the observation of Landau-Zener transitions~\cite{Landau1932PZSLZtransitions,Zener1932PRSLZtransitions}: an initial Fock state can be mapped to two stabilized states of the pumped NEMS when the energy gap between the Fock state and other higher levels becomes small. Due to the energy difference of the stabilized states, the accumulated phase leads to an interference between the two states when mapping the states back to the Fock state basis, exhibiting the sharp oscillations. The evolutions of the energy levels corresponding to these specific cases are shown in Figs.~\ref{fig:3}(b) and \ref{fig:3}(c), respectively.
It is worth noting that the squeezing operation preserves parity, which explains the absence of transitions around $\Delta_{\mathrm{as,KC}}=2K$. These observations are also validated by numerical simulations based on QuTip~\cite{qutip1,qutip2}, showing excellent agreement between the experimental and numerical results. 

The static compensation method shows comparable performance to the dynamic compensation approach when $K<\Delta_{\mathrm{as,KC}}<3K$. This can be attributed to the fact that a precisely symmetric ramping-down process can adiabatically map the stabilized NEMS state back to the original Fock space, regardless of whether the NEMS state is inside or outside the Kerr-cat qubit space. To directly characterize the initialization of Kerr-cat qubits, we perform a direct cat readout~\cite{KCGrimm2020Nature,KCAndy2024arxivDetector,SM} of the NEMS by coherently converting its state to the readout cavity. Measurements along different axes are realized with an X/2 gate implemented through Rabi drives and a Z/2 gate achieved by instantaneously switching off the flux pump~\cite{KCGrimm2020Nature,SM}. 

\begin{figure}[t]
	\centering
	\includegraphics{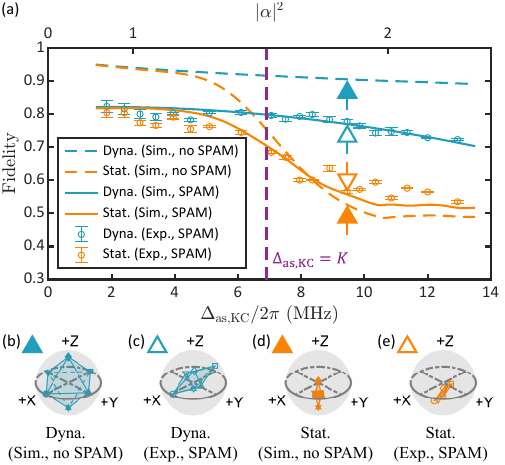}
	\caption{
	(a) Experimental process fidelities of the ramping-up operation for a Kerr-cat qubit, depending on $\Delta_{\mathrm{as,KC}}$ (or $|\alpha|^2$), using different compensation methods. Measurements are performed in the Kerr-cat space.
	The dynamic compensation method performs much better than the static compensation method when the detuning is larger than $K$.
	(b-d) Four example octahedrons on the Bloch sphere. 
	The corresponding locations in (a) are indicated by arrows of different colors, along with solid and hollow arrowheads.
	}
	\label{fig:4}  
\end{figure}

Figure~\ref{fig:4}(a) shows the experimental QPT results (solid markers) for different $\Delta_{\mathrm{as,KC}}$, confirming several key insights into the performance of the dynamic and static methods. When $\Delta_{\mathrm{as,KC}}/K<1$, both methods exhibit comparable initialization fidelities due to the negligible leakage outside the Kerr-cat qubit space. However, as the PIFS increases beyond $K$, the static compensation method experiences a rapid decline in fidelity due to the leakage. In contrast, the dynamic compensation method maintains a high initialization fidelity throughout the entire range of PIFS values. The simulations incorporating the state preparation and measurement (SPAM) errors (solid lines, mainly contributed by the Z/2 gate~\cite{SM}) closely match the experimental data, confirming our understanding of the system dynamics and the source of imperfections. Therefore, based on a theoretical model, we can project the achievable fidelity by excluding SPAM errors (dashed line), demonstrating the robustness and effectiveness of the dynamic compensation approach. In particular, for $\Delta_{\mathrm{as,KC}}/K=1.4$ and $|\alpha|^2=1.9$, the experimental fidelities are 57\% and 78\% for these two approaches, respectively, corresponding to 52\% and 91\% when excluding the SPAM errors. Here, the leakage error of the static compensation method is underestimated due to the imperfections in detecting the leakage states~\cite{KDCGautier2022PRXQ,KCAndy2024arxivDetector}. We find that the fidelity degrades with $\Delta_{\mathrm{as,KC}}$ due to the increasing decoherence rates with the Kerr-cat size, and better performance is possible by improving the coherence of the NEMS. For instance, a Kerr-cat qubit initialization with a fidelity exceeding 99\% is anticipated when the NEMS has $T_\mathrm{1N}=60\,\mathrm{\mu}$s and $T_\mathrm{2N}=30\,\mathrm{\mu}$s.

\textit{Conclusion.-} We have demonstrated high-fidelity initialization of a Kerr-cat qubit using a novel NEMS device with a dynamic compensation method. By carefully analyzing eigenstate evolution of the system under adiabatic pump ramping, we clarify the imperfections arising from PIFS and Landau-Zener transitions. Our experimental and theoretical results consistently show that the dynamic compensation approach outperforms the static compensation scheme, significantly suppressing imperfections associated with large PIFS. At a PIFS to Kerr nonlinearity ratio of $1.4$, the measured initialization fidelity improves from 57\% to 78\%, indicating a projected fidelity of 91\% after excluding SPAM errors. Our demonstration with NEMS and dynamic compensation paves the way for future advancements in scalable Kerr-cat qubits, enabling their application in quantum information processing and exploration of noise-biased fault-tolerant techniques. 

\begin{acknowledgments}
This work was funded by the National Natural Science Foundation of China (Grants No. 11925404, 92165209, 92365301, 12061131011, 92265210, 11890704, 92365206),  Innovation Program for Quantum Science and Technology (Grant No.~2021ZD0300203, 2021ZD0301800, and 2021ZD0300201), and the National Key R\&D Program (2017YFA0304303). This work was also supported by the Fundamental Research Funds for the Central Universities and USTC Research Funds of the Double First-Class Initiative. 
This work was partially carried out at the USTC Center for Micro and Nanoscale Research and Fabrication.
\end{acknowledgments}

\textit{Note added.-} When finalizing this work, we became aware of a related study~\cite{KCIyama2024NCWigner} that also utilized a chirped pulse for initialization. While that study focuses on the characterization and dynamics of cat states, our work delves into the physical mechanism of the adiabatic evolution of the state mapping process.


\begin{thebibliography}{46}%
\makeatletter
\providecommand \@ifxundefined [1]{%
 \@ifx{#1\undefined}
}%
\providecommand \@ifnum [1]{%
 \ifnum #1\expandafter \@firstoftwo
 \else \expandafter \@secondoftwo
 \fi
}%
\providecommand \@ifx [1]{%
 \ifx #1\expandafter \@firstoftwo
 \else \expandafter \@secondoftwo
 \fi
}%
\providecommand \natexlab [1]{#1}%
\providecommand \enquote  [1]{``#1''}%
\providecommand \bibnamefont  [1]{#1}%
\providecommand \bibfnamefont [1]{#1}%
\providecommand \citenamefont [1]{#1}%
\providecommand \href@noop [0]{\@secondoftwo}%
\providecommand \href [0]{\begingroup \@sanitize@url \@href}%
\providecommand \@href[1]{\@@startlink{#1}\@@href}%
\providecommand \@@href[1]{\endgroup#1\@@endlink}%
\providecommand \@sanitize@url [0]{\catcode `\\12\catcode `\$12\catcode
  `\&12\catcode `\#12\catcode `\^12\catcode `\_12\catcode `\%12\relax}%
\providecommand \@@startlink[1]{}%
\providecommand \@@endlink[0]{}%
\providecommand \url  [0]{\begingroup\@sanitize@url \@url }%
\providecommand \@url [1]{\endgroup\@href {#1}{\urlprefix }}%
\providecommand \urlprefix  [0]{URL }%
\providecommand \Eprint [0]{\href }%
\providecommand \doibase [0]{https://doi.org/}%
\providecommand \selectlanguage [0]{\@gobble}%
\providecommand \bibinfo  [0]{\@secondoftwo}%
\providecommand \bibfield  [0]{\@secondoftwo}%
\providecommand \translation [1]{[#1]}%
\providecommand \BibitemOpen [0]{}%
\providecommand \bibitemStop [0]{}%
\providecommand \bibitemNoStop [0]{.\EOS\space}%
\providecommand \EOS [0]{\spacefactor3000\relax}%
\providecommand \BibitemShut  [1]{\csname bibitem#1\endcsname}%
\let\auto@bib@innerbib\@empty
\bibitem [{\citenamefont {Mirrahimi}\ \emph {et~al.}(2014)\citenamefont
  {Mirrahimi}, \citenamefont {Leghtas}, \citenamefont {Albert}, \citenamefont
  {Touzard}, \citenamefont {Schoelkopf}, \citenamefont {Jiang},\ and\
  \citenamefont {Devoret}}]{DCMirrahimi2014NJP}%
  \BibitemOpen
  \bibfield  {author} {\bibinfo {author} {\bibfnamefont {M.}~\bibnamefont
  {Mirrahimi}}, \bibinfo {author} {\bibfnamefont {Z.}~\bibnamefont {Leghtas}},
  \bibinfo {author} {\bibfnamefont {V.~V.}\ \bibnamefont {Albert}}, \bibinfo
  {author} {\bibfnamefont {S.}~\bibnamefont {Touzard}}, \bibinfo {author}
  {\bibfnamefont {R.~J.}\ \bibnamefont {Schoelkopf}}, \bibinfo {author}
  {\bibfnamefont {L.}~\bibnamefont {Jiang}},\ and\ \bibinfo {author}
  {\bibfnamefont {M.~H.}\ \bibnamefont {Devoret}},\ }\href
  {https://iopscience.iop.org/article/10.1088/1367-2630/16/4/045014/meta}
  {\bibfield  {journal} {\bibinfo  {journal} {New Journal of Physics}\ }\textbf
  {\bibinfo {volume} {16}},\ \bibinfo {pages} {045014} (\bibinfo {year}
  {2014})}\BibitemShut {NoStop}%
\bibitem [{\citenamefont {Leghtas}\ \emph {et~al.}(2015)\citenamefont
  {Leghtas}, \citenamefont {Touzard}, \citenamefont {Pop}, \citenamefont {Kou},
  \citenamefont {Vlastakis}, \citenamefont {Petrenko}, \citenamefont {Sliwa},
  \citenamefont {Narla}, \citenamefont {Shankar}, \citenamefont {Hatridge},
  \citenamefont {Reagor}, \citenamefont {Frunzio}, \citenamefont {Schoelkopf},
  \citenamefont {Mirrahimi},\ and\ \citenamefont
  {Devoret}}]{DCZaki2015Science}%
  \BibitemOpen
  \bibfield  {author} {\bibinfo {author} {\bibfnamefont {Z.}~\bibnamefont
  {Leghtas}}, \bibinfo {author} {\bibfnamefont {S.}~\bibnamefont {Touzard}},
  \bibinfo {author} {\bibfnamefont {I.~M.}\ \bibnamefont {Pop}}, \bibinfo
  {author} {\bibfnamefont {A.}~\bibnamefont {Kou}}, \bibinfo {author}
  {\bibfnamefont {B.}~\bibnamefont {Vlastakis}}, \bibinfo {author}
  {\bibfnamefont {A.}~\bibnamefont {Petrenko}}, \bibinfo {author}
  {\bibfnamefont {K.~M.}\ \bibnamefont {Sliwa}}, \bibinfo {author}
  {\bibfnamefont {A.}~\bibnamefont {Narla}}, \bibinfo {author} {\bibfnamefont
  {S.}~\bibnamefont {Shankar}}, \bibinfo {author} {\bibfnamefont {M.~J.}\
  \bibnamefont {Hatridge}}, \bibinfo {author} {\bibfnamefont {M.}~\bibnamefont
  {Reagor}}, \bibinfo {author} {\bibfnamefont {L.}~\bibnamefont {Frunzio}},
  \bibinfo {author} {\bibfnamefont {R.~J.}\ \bibnamefont {Schoelkopf}},
  \bibinfo {author} {\bibfnamefont {M.}~\bibnamefont {Mirrahimi}},\ and\
  \bibinfo {author} {\bibfnamefont {M.~H.}\ \bibnamefont {Devoret}},\ }\href
  {https://doi.org/doi:10.1126/science.aaa2085} {\bibfield  {journal} {\bibinfo
   {journal} {Science}\ }\textbf {\bibinfo {volume} {347}},\ \bibinfo {pages}
  {853} (\bibinfo {year} {2015})}\BibitemShut {NoStop}%
\bibitem [{\citenamefont {Aliferis}\ and\ \citenamefont
  {Preskill}(2008)}]{BPAliferis2008PRAFaultTolerant}%
  \BibitemOpen
  \bibfield  {author} {\bibinfo {author} {\bibfnamefont {P.}~\bibnamefont
  {Aliferis}}\ and\ \bibinfo {author} {\bibfnamefont {J.}~\bibnamefont
  {Preskill}},\ }\href {https://doi.org/10.1103/PhysRevA.78.052331} {\bibfield
  {journal} {\bibinfo  {journal} {Physical Review A}\ }\textbf {\bibinfo
  {volume} {78}},\ \bibinfo {pages} {052331} (\bibinfo {year}
  {2008})}\BibitemShut {NoStop}%
\bibitem [{\citenamefont {Aliferis}\ \emph {et~al.}(2009)\citenamefont
  {Aliferis}, \citenamefont {Brito}, \citenamefont {DiVincenzo}, \citenamefont
  {Preskill}, \citenamefont {Steffen},\ and\ \citenamefont
  {Terhal}}]{BPAliferis2009NJPFaultTolerantCase}%
  \BibitemOpen
  \bibfield  {author} {\bibinfo {author} {\bibfnamefont {P.}~\bibnamefont
  {Aliferis}}, \bibinfo {author} {\bibfnamefont {F.}~\bibnamefont {Brito}},
  \bibinfo {author} {\bibfnamefont {D.~P.}\ \bibnamefont {DiVincenzo}},
  \bibinfo {author} {\bibfnamefont {J.}~\bibnamefont {Preskill}}, \bibinfo
  {author} {\bibfnamefont {M.}~\bibnamefont {Steffen}},\ and\ \bibinfo {author}
  {\bibfnamefont {B.~M.}\ \bibnamefont {Terhal}},\ }\href
  {https://doi.org/10.1088/1367-2630/11/1/013061} {\bibfield  {journal}
  {\bibinfo  {journal} {New Journal of Physics}\ }\textbf {\bibinfo {volume}
  {11}},\ \bibinfo {pages} {013061} (\bibinfo {year} {2009})}\BibitemShut
  {NoStop}%
\bibitem [{\citenamefont {Webster}\ \emph {et~al.}(2015)\citenamefont
  {Webster}, \citenamefont {Bartlett},\ and\ \citenamefont
  {Poulin}}]{BPWebster2015PRA}%
  \BibitemOpen
  \bibfield  {author} {\bibinfo {author} {\bibfnamefont {P.}~\bibnamefont
  {Webster}}, \bibinfo {author} {\bibfnamefont {S.~D.}\ \bibnamefont
  {Bartlett}},\ and\ \bibinfo {author} {\bibfnamefont {D.}~\bibnamefont
  {Poulin}},\ }\href {https://doi.org/10.1103/PhysRevA.92.062309} {\bibfield
  {journal} {\bibinfo  {journal} {Physical Review A}\ }\textbf {\bibinfo
  {volume} {92}},\ \bibinfo {pages} {062309} (\bibinfo {year}
  {2015})}\BibitemShut {NoStop}%
\bibitem [{\citenamefont {Chamberland}\ \emph {et~al.}(2022)\citenamefont
  {Chamberland}, \citenamefont {Noh}, \citenamefont {Arrangoiz-Arriola},
  \citenamefont {Campbell}, \citenamefont {Hann}, \citenamefont {Iverson},
  \citenamefont {Putterman}, \citenamefont {Bohdanowicz}, \citenamefont
  {Flammia}, \citenamefont {Keller}, \citenamefont {Refael}, \citenamefont
  {Preskill}, \citenamefont {Jiang}, \citenamefont {Safavi-Naeini},
  \citenamefont {Painter},\ and\ \citenamefont
  {Brand\~{a}o}}]{BPChamberland2022PRXQFaultToleDissCat}%
  \BibitemOpen
  \bibfield  {author} {\bibinfo {author} {\bibfnamefont {C.}~\bibnamefont
  {Chamberland}}, \bibinfo {author} {\bibfnamefont {K.}~\bibnamefont {Noh}},
  \bibinfo {author} {\bibfnamefont {P.}~\bibnamefont {Arrangoiz-Arriola}},
  \bibinfo {author} {\bibfnamefont {E.~T.}\ \bibnamefont {Campbell}}, \bibinfo
  {author} {\bibfnamefont {C.~T.}\ \bibnamefont {Hann}}, \bibinfo {author}
  {\bibfnamefont {J.}~\bibnamefont {Iverson}}, \bibinfo {author} {\bibfnamefont
  {H.}~\bibnamefont {Putterman}}, \bibinfo {author} {\bibfnamefont {T.~C.}\
  \bibnamefont {Bohdanowicz}}, \bibinfo {author} {\bibfnamefont {S.~T.}\
  \bibnamefont {Flammia}}, \bibinfo {author} {\bibfnamefont {A.}~\bibnamefont
  {Keller}}, \bibinfo {author} {\bibfnamefont {G.}~\bibnamefont {Refael}},
  \bibinfo {author} {\bibfnamefont {J.}~\bibnamefont {Preskill}}, \bibinfo
  {author} {\bibfnamefont {L.}~\bibnamefont {Jiang}}, \bibinfo {author}
  {\bibfnamefont {A.~H.}\ \bibnamefont {Safavi-Naeini}}, \bibinfo {author}
  {\bibfnamefont {O.}~\bibnamefont {Painter}},\ and\ \bibinfo {author}
  {\bibfnamefont {F.~G. S.~L.}\ \bibnamefont {Brand\~{a}o}},\ }\href
  {https://doi.org/10.1103/PRXQuantum.3.010329} {\bibfield  {journal} {\bibinfo
   {journal} {PRX Quantum}\ }\textbf {\bibinfo {volume} {3}},\ \bibinfo {pages}
  {010329} (\bibinfo {year} {2022})}\BibitemShut {NoStop}%
\bibitem [{\citenamefont {Lescanne}\ \emph {et~al.}(2020)\citenamefont
  {Lescanne}, \citenamefont {Villiers}, \citenamefont {Peronnin}, \citenamefont
  {Sarlette}, \citenamefont {Delbecq}, \citenamefont {Huard}, \citenamefont
  {Kontos}, \citenamefont {Mirrahimi},\ and\ \citenamefont
  {Leghtas}}]{DCLescanne2020NP}%
  \BibitemOpen
  \bibfield  {author} {\bibinfo {author} {\bibfnamefont {R.}~\bibnamefont
  {Lescanne}}, \bibinfo {author} {\bibfnamefont {M.}~\bibnamefont {Villiers}},
  \bibinfo {author} {\bibfnamefont {T.}~\bibnamefont {Peronnin}}, \bibinfo
  {author} {\bibfnamefont {A.}~\bibnamefont {Sarlette}}, \bibinfo {author}
  {\bibfnamefont {M.}~\bibnamefont {Delbecq}}, \bibinfo {author} {\bibfnamefont
  {B.}~\bibnamefont {Huard}}, \bibinfo {author} {\bibfnamefont
  {T.}~\bibnamefont {Kontos}}, \bibinfo {author} {\bibfnamefont
  {M.}~\bibnamefont {Mirrahimi}},\ and\ \bibinfo {author} {\bibfnamefont
  {Z.}~\bibnamefont {Leghtas}},\ }\href
  {https://doi.org/10.1038/s41567-020-0824-x} {\bibfield  {journal} {\bibinfo
  {journal} {Nature Physics}\ }\textbf {\bibinfo {volume} {16}},\ \bibinfo
  {pages} {509} (\bibinfo {year} {2020})}\BibitemShut {NoStop}%
\bibitem [{\citenamefont {Puri}\ \emph {et~al.}(2017)\citenamefont {Puri},
  \citenamefont {Boutin},\ and\ \citenamefont {Blais}}]{KCPuri2017npjQI}%
  \BibitemOpen
  \bibfield  {author} {\bibinfo {author} {\bibfnamefont {S.}~\bibnamefont
  {Puri}}, \bibinfo {author} {\bibfnamefont {S.}~\bibnamefont {Boutin}},\ and\
  \bibinfo {author} {\bibfnamefont {A.}~\bibnamefont {Blais}},\ }\href
  {https://doi.org/10.1038/s41534-017-0019-1} {\bibfield  {journal} {\bibinfo
  {journal} {npj Quantum Information}\ }\textbf {\bibinfo {volume} {3}},\
  \bibinfo {pages} {18} (\bibinfo {year} {2017})}\BibitemShut {NoStop}%
\bibitem [{\citenamefont {Fowler}\ \emph {et~al.}(2012)\citenamefont {Fowler},
  \citenamefont {Whiteside},\ and\ \citenamefont
  {Hollenberg}}]{SurfaceCodeFowler2012PRL}%
  \BibitemOpen
  \bibfield  {author} {\bibinfo {author} {\bibfnamefont {A.~G.}\ \bibnamefont
  {Fowler}}, \bibinfo {author} {\bibfnamefont {A.~C.}\ \bibnamefont
  {Whiteside}},\ and\ \bibinfo {author} {\bibfnamefont {L.~C.~L.}\ \bibnamefont
  {Hollenberg}},\ }\href {https://doi.org/10.1103/PhysRevLett.108.180501}
  {\bibfield  {journal} {\bibinfo  {journal} {Physical Review Letters}\
  }\textbf {\bibinfo {volume} {108}},\ \bibinfo {pages} {180501} (\bibinfo
  {year} {2012})}\BibitemShut {NoStop}%
\bibitem [{\citenamefont {Bonilla~Ataides}\ \emph {et~al.}(2021)\citenamefont
  {Bonilla~Ataides}, \citenamefont {Tuckett}, \citenamefont {Bartlett},
  \citenamefont {Flammia},\ and\ \citenamefont
  {Brown}}]{XZZXCodeAtaides2021NC}%
  \BibitemOpen
  \bibfield  {author} {\bibinfo {author} {\bibfnamefont {J.~P.}\ \bibnamefont
  {Bonilla~Ataides}}, \bibinfo {author} {\bibfnamefont {D.~K.}\ \bibnamefont
  {Tuckett}}, \bibinfo {author} {\bibfnamefont {S.~D.}\ \bibnamefont
  {Bartlett}}, \bibinfo {author} {\bibfnamefont {S.~T.}\ \bibnamefont
  {Flammia}},\ and\ \bibinfo {author} {\bibfnamefont {B.~J.}\ \bibnamefont
  {Brown}},\ }\href {https://doi.org/10.1038/s41467-021-22274-1} {\bibfield
  {journal} {\bibinfo  {journal} {Nature Communications}\ }\textbf {\bibinfo
  {volume} {12}},\ \bibinfo {pages} {2172} (\bibinfo {year}
  {2021})}\BibitemShut {NoStop}%
\bibitem [{\citenamefont {Tuckett}\ \emph {et~al.}(2018)\citenamefont
  {Tuckett}, \citenamefont {Bartlett},\ and\ \citenamefont
  {Flammia}}]{BPTuckett2018PRLBiasQubitSurface}%
  \BibitemOpen
  \bibfield  {author} {\bibinfo {author} {\bibfnamefont {D.~K.}\ \bibnamefont
  {Tuckett}}, \bibinfo {author} {\bibfnamefont {S.~D.}\ \bibnamefont
  {Bartlett}},\ and\ \bibinfo {author} {\bibfnamefont {S.~T.}\ \bibnamefont
  {Flammia}},\ }\href {https://doi.org/10.1103/PhysRevLett.120.050505}
  {\bibfield  {journal} {\bibinfo  {journal} {Physical Review Letters}\
  }\textbf {\bibinfo {volume} {120}},\ \bibinfo {pages} {050505} (\bibinfo
  {year} {2018})}\BibitemShut {NoStop}%
\bibitem [{\citenamefont {Tuckett}\ \emph {et~al.}(2019)\citenamefont
  {Tuckett}, \citenamefont {Darmawan}, \citenamefont {Chubb}, \citenamefont
  {Bravyi}, \citenamefont {Bartlett},\ and\ \citenamefont
  {Flammia}}]{BPTuckett2019PRXBiasQubitSurface}%
  \BibitemOpen
  \bibfield  {author} {\bibinfo {author} {\bibfnamefont {D.~K.}\ \bibnamefont
  {Tuckett}}, \bibinfo {author} {\bibfnamefont {A.~S.}\ \bibnamefont
  {Darmawan}}, \bibinfo {author} {\bibfnamefont {C.~T.}\ \bibnamefont {Chubb}},
  \bibinfo {author} {\bibfnamefont {S.}~\bibnamefont {Bravyi}}, \bibinfo
  {author} {\bibfnamefont {S.~D.}\ \bibnamefont {Bartlett}},\ and\ \bibinfo
  {author} {\bibfnamefont {S.~T.}\ \bibnamefont {Flammia}},\ }\href
  {https://doi.org/10.1103/PhysRevX.9.041031} {\bibfield  {journal} {\bibinfo
  {journal} {Physical Review X}\ }\textbf {\bibinfo {volume} {9}},\ \bibinfo
  {pages} {041031} (\bibinfo {year} {2019})}\BibitemShut {NoStop}%
\bibitem [{\citenamefont {Tuckett}\ \emph {et~al.}(2020)\citenamefont
  {Tuckett}, \citenamefont {Bartlett}, \citenamefont {Flammia},\ and\
  \citenamefont {Brown}}]{BPTuckett2020PRLBiasQubitSurface}%
  \BibitemOpen
  \bibfield  {author} {\bibinfo {author} {\bibfnamefont {D.~K.}\ \bibnamefont
  {Tuckett}}, \bibinfo {author} {\bibfnamefont {S.~D.}\ \bibnamefont
  {Bartlett}}, \bibinfo {author} {\bibfnamefont {S.~T.}\ \bibnamefont
  {Flammia}},\ and\ \bibinfo {author} {\bibfnamefont {B.~J.}\ \bibnamefont
  {Brown}},\ }\href {https://doi.org/10.1103/PhysRevLett.124.130501} {\bibfield
   {journal} {\bibinfo  {journal} {Physical Review Letters}\ }\textbf {\bibinfo
  {volume} {124}},\ \bibinfo {pages} {130501} (\bibinfo {year}
  {2020})}\BibitemShut {NoStop}%
\bibitem [{\citenamefont {Darmawan}\ \emph {et~al.}(2021)\citenamefont
  {Darmawan}, \citenamefont {Brown}, \citenamefont {Grimsmo}, \citenamefont
  {Tuckett},\ and\ \citenamefont {Puri}}]{BPDarmawan2021PRXQXZZXKerrCat}%
  \BibitemOpen
  \bibfield  {author} {\bibinfo {author} {\bibfnamefont {A.~S.}\ \bibnamefont
  {Darmawan}}, \bibinfo {author} {\bibfnamefont {B.~J.}\ \bibnamefont {Brown}},
  \bibinfo {author} {\bibfnamefont {A.~L.}\ \bibnamefont {Grimsmo}}, \bibinfo
  {author} {\bibfnamefont {D.~K.}\ \bibnamefont {Tuckett}},\ and\ \bibinfo
  {author} {\bibfnamefont {S.}~\bibnamefont {Puri}},\ }\href
  {https://doi.org/10.1103/PRXQuantum.2.030345} {\bibfield  {journal} {\bibinfo
   {journal} {PRX Quantum}\ }\textbf {\bibinfo {volume} {2}},\ \bibinfo {pages}
  {030345} (\bibinfo {year} {2021})}\BibitemShut {NoStop}%
\bibitem [{\citenamefont {Shor}(1995)}]{Shor1995PRAcode}%
  \BibitemOpen
  \bibfield  {author} {\bibinfo {author} {\bibfnamefont {P.~W.}\ \bibnamefont
  {Shor}},\ }\href {https://doi.org/10.1103/PhysRevA.52.R2493} {\bibfield
  {journal} {\bibinfo  {journal} {Physical Review A}\ }\textbf {\bibinfo
  {volume} {52}},\ \bibinfo {pages} {R2493} (\bibinfo {year}
  {1995})}\BibitemShut {NoStop}%
\bibitem [{\citenamefont {Steane}(1996)}]{Steane1996PRLcode}%
  \BibitemOpen
  \bibfield  {author} {\bibinfo {author} {\bibfnamefont {A.~M.}\ \bibnamefont
  {Steane}},\ }\href {https://doi.org/10.1103/PhysRevLett.77.793} {\bibfield
  {journal} {\bibinfo  {journal} {Physical Review Letters}\ }\textbf {\bibinfo
  {volume} {77}},\ \bibinfo {pages} {793} (\bibinfo {year} {1996})}\BibitemShut
  {NoStop}%
\bibitem [{\citenamefont {Guillaud}\ and\ \citenamefont
  {Mirrahimi}(2019)}]{DCGuillaud2019PRXRepetition}%
  \BibitemOpen
  \bibfield  {author} {\bibinfo {author} {\bibfnamefont {J.}~\bibnamefont
  {Guillaud}}\ and\ \bibinfo {author} {\bibfnamefont {M.}~\bibnamefont
  {Mirrahimi}},\ }\href {https://doi.org/10.1103/PhysRevX.9.041053} {\bibfield
  {journal} {\bibinfo  {journal} {Physical Review X}\ }\textbf {\bibinfo
  {volume} {9}},\ \bibinfo {pages} {041053} (\bibinfo {year}
  {2019})}\BibitemShut {NoStop}%
\bibitem [{\citenamefont {Guillaud}\ and\ \citenamefont
  {Mirrahimi}(2021)}]{DCGuillaud2021PRARepetitionThreshold}%
  \BibitemOpen
  \bibfield  {author} {\bibinfo {author} {\bibfnamefont {J.}~\bibnamefont
  {Guillaud}}\ and\ \bibinfo {author} {\bibfnamefont {M.}~\bibnamefont
  {Mirrahimi}},\ }\href {https://doi.org/10.1103/PhysRevA.103.042413}
  {\bibfield  {journal} {\bibinfo  {journal} {Physical Review A}\ }\textbf
  {\bibinfo {volume} {103}},\ \bibinfo {pages} {042413} (\bibinfo {year}
  {2021})}\BibitemShut {NoStop}%
\bibitem [{\citenamefont {Gouzien}\ \emph {et~al.}(2023)\citenamefont
  {Gouzien}, \citenamefont {Ruiz}, \citenamefont {Le~R\'{e}gent}, \citenamefont
  {Guillaud},\ and\ \citenamefont {Sangouard}}]{DCGouzien2023PRLRepetitiond5}%
  \BibitemOpen
  \bibfield  {author} {\bibinfo {author} {\bibfnamefont {E.}~\bibnamefont
  {Gouzien}}, \bibinfo {author} {\bibfnamefont {D.}~\bibnamefont {Ruiz}},
  \bibinfo {author} {\bibfnamefont {F.-M.}\ \bibnamefont {Le~R\'{e}gent}},
  \bibinfo {author} {\bibfnamefont {J.}~\bibnamefont {Guillaud}},\ and\
  \bibinfo {author} {\bibfnamefont {N.}~\bibnamefont {Sangouard}},\ }\href
  {https://doi.org/10.1103/PhysRevLett.131.040602} {\bibfield  {journal}
  {\bibinfo  {journal} {Physical Review Letters}\ }\textbf {\bibinfo {volume}
  {131}},\ \bibinfo {pages} {040602} (\bibinfo {year} {2023})}\BibitemShut
  {NoStop}%
\bibitem [{\citenamefont {Puri}\ \emph {et~al.}(2019)\citenamefont {Puri},
  \citenamefont {Grimm}, \citenamefont {Campagne-Ibarcq}, \citenamefont
  {Eickbusch}, \citenamefont {Noh}, \citenamefont {Roberts}, \citenamefont
  {Jiang}, \citenamefont {Mirrahimi}, \citenamefont {Devoret},\ and\
  \citenamefont {Girvin}}]{KCPuri2019PRXDetector}%
  \BibitemOpen
  \bibfield  {author} {\bibinfo {author} {\bibfnamefont {S.}~\bibnamefont
  {Puri}}, \bibinfo {author} {\bibfnamefont {A.}~\bibnamefont {Grimm}},
  \bibinfo {author} {\bibfnamefont {P.}~\bibnamefont {Campagne-Ibarcq}},
  \bibinfo {author} {\bibfnamefont {A.}~\bibnamefont {Eickbusch}}, \bibinfo
  {author} {\bibfnamefont {K.}~\bibnamefont {Noh}}, \bibinfo {author}
  {\bibfnamefont {G.}~\bibnamefont {Roberts}}, \bibinfo {author} {\bibfnamefont
  {L.}~\bibnamefont {Jiang}}, \bibinfo {author} {\bibfnamefont
  {M.}~\bibnamefont {Mirrahimi}}, \bibinfo {author} {\bibfnamefont {M.~H.}\
  \bibnamefont {Devoret}},\ and\ \bibinfo {author} {\bibfnamefont {S.~M.}\
  \bibnamefont {Girvin}},\ }\href {https://doi.org/10.1103/PhysRevX.9.041009}
  {\bibfield  {journal} {\bibinfo  {journal} {Physical Review X}\ }\textbf
  {\bibinfo {volume} {9}},\ \bibinfo {pages} {041009} (\bibinfo {year}
  {2019})}\BibitemShut {NoStop}%
\bibitem [{\citenamefont {Ding}\ \emph {et~al.}(2024)\citenamefont {Ding},
  \citenamefont {Brock}, \citenamefont {Eickbusch}, \citenamefont
  {Koottandavida}, \citenamefont {Frattini}, \citenamefont {Cortinas},
  \citenamefont {Joshi}, \citenamefont {de~Graaf}, \citenamefont {Chapman},
  \citenamefont {Ganjam}, \citenamefont {Frunzio}, \citenamefont {Schoelkopf},\
  and\ \citenamefont {Devoret}}]{KCAndy2024arxivDetector}%
  \BibitemOpen
  \bibfield  {author} {\bibinfo {author} {\bibfnamefont {A.~Z.}\ \bibnamefont
  {Ding}}, \bibinfo {author} {\bibfnamefont {B.~L.}\ \bibnamefont {Brock}},
  \bibinfo {author} {\bibfnamefont {A.}~\bibnamefont {Eickbusch}}, \bibinfo
  {author} {\bibfnamefont {A.}~\bibnamefont {Koottandavida}}, \bibinfo {author}
  {\bibfnamefont {N.~E.}\ \bibnamefont {Frattini}}, \bibinfo {author}
  {\bibfnamefont {R.~G.}\ \bibnamefont {Cortinas}}, \bibinfo {author}
  {\bibfnamefont {V.~R.}\ \bibnamefont {Joshi}}, \bibinfo {author}
  {\bibfnamefont {S.~J.}\ \bibnamefont {de~Graaf}}, \bibinfo {author}
  {\bibfnamefont {B.~J.}\ \bibnamefont {Chapman}}, \bibinfo {author}
  {\bibfnamefont {S.}~\bibnamefont {Ganjam}}, \bibinfo {author} {\bibfnamefont
  {L.}~\bibnamefont {Frunzio}}, \bibinfo {author} {\bibfnamefont {R.~J.}\
  \bibnamefont {Schoelkopf}},\ and\ \bibinfo {author} {\bibfnamefont {M.~H.}\
  \bibnamefont {Devoret}},\ }\href {https://arxiv.org/abs/2407.10940}
  {\bibfield  {journal} {\bibinfo  {journal} {arXiv:2407.10940}\ } (\bibinfo
  {year} {2024})}\BibitemShut {NoStop}%
\bibitem [{\citenamefont {Berdou}\ \emph {et~al.}(2023)\citenamefont {Berdou},
  \citenamefont {Murani}, \citenamefont {R\'{e}glade}, \citenamefont {Smith},
  \citenamefont {Villiers}, \citenamefont {Palomo}, \citenamefont {Rosticher},
  \citenamefont {Denis}, \citenamefont {Morfin}, \citenamefont {Delbecq},
  \citenamefont {Kontos}, \citenamefont {Pankratova}, \citenamefont
  {Rautschke}, \citenamefont {Peronnin}, \citenamefont {Sellem}, \citenamefont
  {Rouchon}, \citenamefont {Sarlette}, \citenamefont {Mirrahimi}, \citenamefont
  {Campagne-Ibarcq}, \citenamefont {Jezouin}, \citenamefont {Lescanne},\ and\
  \citenamefont {Leghtas}}]{DCBerdou2023PRXQ100s}%
  \BibitemOpen
  \bibfield  {author} {\bibinfo {author} {\bibfnamefont {C.}~\bibnamefont
  {Berdou}}, \bibinfo {author} {\bibfnamefont {A.}~\bibnamefont {Murani}},
  \bibinfo {author} {\bibfnamefont {U.}~\bibnamefont {R\'{e}glade}}, \bibinfo
  {author} {\bibfnamefont {W.~C.}\ \bibnamefont {Smith}}, \bibinfo {author}
  {\bibfnamefont {M.}~\bibnamefont {Villiers}}, \bibinfo {author}
  {\bibfnamefont {J.}~\bibnamefont {Palomo}}, \bibinfo {author} {\bibfnamefont
  {M.}~\bibnamefont {Rosticher}}, \bibinfo {author} {\bibfnamefont
  {A.}~\bibnamefont {Denis}}, \bibinfo {author} {\bibfnamefont
  {P.}~\bibnamefont {Morfin}}, \bibinfo {author} {\bibfnamefont
  {M.}~\bibnamefont {Delbecq}}, \bibinfo {author} {\bibfnamefont
  {T.}~\bibnamefont {Kontos}}, \bibinfo {author} {\bibfnamefont
  {N.}~\bibnamefont {Pankratova}}, \bibinfo {author} {\bibfnamefont
  {F.}~\bibnamefont {Rautschke}}, \bibinfo {author} {\bibfnamefont
  {T.}~\bibnamefont {Peronnin}}, \bibinfo {author} {\bibfnamefont {L.~A.}\
  \bibnamefont {Sellem}}, \bibinfo {author} {\bibfnamefont {P.}~\bibnamefont
  {Rouchon}}, \bibinfo {author} {\bibfnamefont {A.}~\bibnamefont {Sarlette}},
  \bibinfo {author} {\bibfnamefont {M.}~\bibnamefont {Mirrahimi}}, \bibinfo
  {author} {\bibfnamefont {P.}~\bibnamefont {Campagne-Ibarcq}}, \bibinfo
  {author} {\bibfnamefont {S.}~\bibnamefont {Jezouin}}, \bibinfo {author}
  {\bibfnamefont {R.}~\bibnamefont {Lescanne}},\ and\ \bibinfo {author}
  {\bibfnamefont {Z.}~\bibnamefont {Leghtas}},\ }\href
  {https://doi.org/10.1103/PRXQuantum.4.020350} {\bibfield  {journal} {\bibinfo
   {journal} {PRX Quantum}\ }\textbf {\bibinfo {volume} {4}},\ \bibinfo {pages}
  {020350} (\bibinfo {year} {2023})}\BibitemShut {NoStop}%
\bibitem [{\citenamefont {R\'{e}glade}\ \emph {et~al.}(2024)\citenamefont
  {R\'{e}glade}, \citenamefont {Bocquet}, \citenamefont {Gautier},
  \citenamefont {Cohen}, \citenamefont {Marquet}, \citenamefont {Albertinale},
  \citenamefont {Pankratova}, \citenamefont {Hall\'{e}n}, \citenamefont
  {Rautschke}, \citenamefont {Sellem}, \citenamefont {Rouchon}, \citenamefont
  {Sarlette}, \citenamefont {Mirrahimi}, \citenamefont {Campagne-Ibarcq},
  \citenamefont {Lescanne}, \citenamefont {Jezouin},\ and\ \citenamefont
  {Leghtas}}]{DCReglade2024Nature10sControl}%
  \BibitemOpen
  \bibfield  {author} {\bibinfo {author} {\bibfnamefont {U.}~\bibnamefont
  {R\'{e}glade}}, \bibinfo {author} {\bibfnamefont {A.}~\bibnamefont
  {Bocquet}}, \bibinfo {author} {\bibfnamefont {R.}~\bibnamefont {Gautier}},
  \bibinfo {author} {\bibfnamefont {J.}~\bibnamefont {Cohen}}, \bibinfo
  {author} {\bibfnamefont {A.}~\bibnamefont {Marquet}}, \bibinfo {author}
  {\bibfnamefont {E.}~\bibnamefont {Albertinale}}, \bibinfo {author}
  {\bibfnamefont {N.}~\bibnamefont {Pankratova}}, \bibinfo {author}
  {\bibfnamefont {M.}~\bibnamefont {Hall\'{e}n}}, \bibinfo {author}
  {\bibfnamefont {F.}~\bibnamefont {Rautschke}}, \bibinfo {author}
  {\bibfnamefont {L.~A.}\ \bibnamefont {Sellem}}, \bibinfo {author}
  {\bibfnamefont {P.}~\bibnamefont {Rouchon}}, \bibinfo {author} {\bibfnamefont
  {A.}~\bibnamefont {Sarlette}}, \bibinfo {author} {\bibfnamefont
  {M.}~\bibnamefont {Mirrahimi}}, \bibinfo {author} {\bibfnamefont
  {P.}~\bibnamefont {Campagne-Ibarcq}}, \bibinfo {author} {\bibfnamefont
  {R.}~\bibnamefont {Lescanne}}, \bibinfo {author} {\bibfnamefont
  {S.}~\bibnamefont {Jezouin}},\ and\ \bibinfo {author} {\bibfnamefont
  {Z.}~\bibnamefont {Leghtas}},\ }\href
  {https://doi.org/10.1038/s41586-024-07294-3} {\bibfield  {journal} {\bibinfo
  {journal} {Nature}\ }\textbf {\bibinfo {volume} {629}},\ \bibinfo {pages}
  {778} (\bibinfo {year} {2024})}\BibitemShut {NoStop}%
\bibitem [{\citenamefont {Grimm}\ \emph {et~al.}(2020)\citenamefont {Grimm},
  \citenamefont {Frattini}, \citenamefont {Puri}, \citenamefont {Mundhada},
  \citenamefont {Touzard}, \citenamefont {Mirrahimi}, \citenamefont {Girvin},
  \citenamefont {Shankar},\ and\ \citenamefont {Devoret}}]{KCGrimm2020Nature}%
  \BibitemOpen
  \bibfield  {author} {\bibinfo {author} {\bibfnamefont {A.}~\bibnamefont
  {Grimm}}, \bibinfo {author} {\bibfnamefont {N.~E.}\ \bibnamefont {Frattini}},
  \bibinfo {author} {\bibfnamefont {S.}~\bibnamefont {Puri}}, \bibinfo {author}
  {\bibfnamefont {S.~O.}\ \bibnamefont {Mundhada}}, \bibinfo {author}
  {\bibfnamefont {S.}~\bibnamefont {Touzard}}, \bibinfo {author} {\bibfnamefont
  {M.}~\bibnamefont {Mirrahimi}}, \bibinfo {author} {\bibfnamefont {S.~M.}\
  \bibnamefont {Girvin}}, \bibinfo {author} {\bibfnamefont {S.}~\bibnamefont
  {Shankar}},\ and\ \bibinfo {author} {\bibfnamefont {M.~H.}\ \bibnamefont
  {Devoret}},\ }\href {https://doi.org/10.1038/s41586-020-2587-z} {\bibfield
  {journal} {\bibinfo  {journal} {Nature}\ }\textbf {\bibinfo {volume} {584}},\
  \bibinfo {pages} {205} (\bibinfo {year} {2020})}\BibitemShut {NoStop}%
\bibitem [{\citenamefont {Xu}\ \emph {et~al.}(2022)\citenamefont {Xu},
  \citenamefont {Iverson}, \citenamefont {Brand\~{a}o},\ and\ \citenamefont
  {Jiang}}]{KCXu2022PRRKCatFast}%
  \BibitemOpen
  \bibfield  {author} {\bibinfo {author} {\bibfnamefont {Q.}~\bibnamefont
  {Xu}}, \bibinfo {author} {\bibfnamefont {J.~K.}\ \bibnamefont {Iverson}},
  \bibinfo {author} {\bibfnamefont {F.~G. S.~L.}\ \bibnamefont {Brand\~{a}o}},\
  and\ \bibinfo {author} {\bibfnamefont {L.}~\bibnamefont {Jiang}},\ }\href
  {https://doi.org/10.1103/PhysRevResearch.4.013082} {\bibfield  {journal}
  {\bibinfo  {journal} {Physical Review Research}\ }\textbf {\bibinfo {volume}
  {4}},\ \bibinfo {pages} {013082} (\bibinfo {year} {2022})}\BibitemShut
  {NoStop}%
\bibitem [{\citenamefont {Lu}\ \emph {et~al.}(2023)\citenamefont {Lu},
  \citenamefont {Maiti}, \citenamefont {Garmon}, \citenamefont {Ganjam},
  \citenamefont {Zhang}, \citenamefont {Claes}, \citenamefont {Frunzio},
  \citenamefont {Girvin},\ and\ \citenamefont
  {Schoelkopf}}]{FPLu2023NCBeamSplitting}%
  \BibitemOpen
  \bibfield  {author} {\bibinfo {author} {\bibfnamefont {Y.}~\bibnamefont
  {Lu}}, \bibinfo {author} {\bibfnamefont {A.}~\bibnamefont {Maiti}}, \bibinfo
  {author} {\bibfnamefont {J.~W.~O.}\ \bibnamefont {Garmon}}, \bibinfo {author}
  {\bibfnamefont {S.}~\bibnamefont {Ganjam}}, \bibinfo {author} {\bibfnamefont
  {Y.}~\bibnamefont {Zhang}}, \bibinfo {author} {\bibfnamefont
  {J.}~\bibnamefont {Claes}}, \bibinfo {author} {\bibfnamefont
  {L.}~\bibnamefont {Frunzio}}, \bibinfo {author} {\bibfnamefont {S.~M.}\
  \bibnamefont {Girvin}},\ and\ \bibinfo {author} {\bibfnamefont {R.~J.}\
  \bibnamefont {Schoelkopf}},\ }\href
  {https://doi.org/10.1038/s41467-023-41104-0} {\bibfield  {journal} {\bibinfo
  {journal} {Nature Communications}\ }\textbf {\bibinfo {volume} {14}},\
  \bibinfo {pages} {5767} (\bibinfo {year} {2023})}\BibitemShut {NoStop}%
\bibitem [{\citenamefont {Iyama}\ \emph {et~al.}(2024)\citenamefont {Iyama},
  \citenamefont {Kamiya}, \citenamefont {Fujii}, \citenamefont {Mukai},
  \citenamefont {Zhou}, \citenamefont {Nagase}, \citenamefont {Tomonaga},
  \citenamefont {Wang}, \citenamefont {Xue}, \citenamefont {Watabe},
  \citenamefont {Kwon},\ and\ \citenamefont {Tsai}}]{KCIyama2024NCWigner}%
  \BibitemOpen
  \bibfield  {author} {\bibinfo {author} {\bibfnamefont {D.}~\bibnamefont
  {Iyama}}, \bibinfo {author} {\bibfnamefont {T.}~\bibnamefont {Kamiya}},
  \bibinfo {author} {\bibfnamefont {S.}~\bibnamefont {Fujii}}, \bibinfo
  {author} {\bibfnamefont {H.}~\bibnamefont {Mukai}}, \bibinfo {author}
  {\bibfnamefont {Y.}~\bibnamefont {Zhou}}, \bibinfo {author} {\bibfnamefont
  {T.}~\bibnamefont {Nagase}}, \bibinfo {author} {\bibfnamefont
  {A.}~\bibnamefont {Tomonaga}}, \bibinfo {author} {\bibfnamefont
  {R.}~\bibnamefont {Wang}}, \bibinfo {author} {\bibfnamefont {J.-J.}\
  \bibnamefont {Xue}}, \bibinfo {author} {\bibfnamefont {S.}~\bibnamefont
  {Watabe}}, \bibinfo {author} {\bibfnamefont {S.}~\bibnamefont {Kwon}},\ and\
  \bibinfo {author} {\bibfnamefont {J.-S.}\ \bibnamefont {Tsai}},\ }\href
  {https://doi.org/10.1038/s41467-023-44496-1} {\bibfield  {journal} {\bibinfo
  {journal} {Nature Communications}\ }\textbf {\bibinfo {volume} {15}},\
  \bibinfo {pages} {86} (\bibinfo {year} {2024})}\BibitemShut {NoStop}%
\bibitem [{\citenamefont {Hoshi}\ \emph {et~al.}(2024)\citenamefont {Hoshi},
  \citenamefont {Nagase}, \citenamefont {Kwon}, \citenamefont {Iyama},
  \citenamefont {Kamiya}, \citenamefont {Fujii}, \citenamefont {Mukai},
  \citenamefont {Ahmed}, \citenamefont {Kockum}, \citenamefont {Watabe},
  \citenamefont {Yoshihara},\ and\ \citenamefont {Tsai}}]{KCHoshiArxiv2KPO}%
  \BibitemOpen
  \bibfield  {author} {\bibinfo {author} {\bibfnamefont {D.}~\bibnamefont
  {Hoshi}}, \bibinfo {author} {\bibfnamefont {T.}~\bibnamefont {Nagase}},
  \bibinfo {author} {\bibfnamefont {S.}~\bibnamefont {Kwon}}, \bibinfo {author}
  {\bibfnamefont {D.}~\bibnamefont {Iyama}}, \bibinfo {author} {\bibfnamefont
  {T.}~\bibnamefont {Kamiya}}, \bibinfo {author} {\bibfnamefont
  {S.}~\bibnamefont {Fujii}}, \bibinfo {author} {\bibfnamefont
  {H.}~\bibnamefont {Mukai}}, \bibinfo {author} {\bibfnamefont
  {S.}~\bibnamefont {Ahmed}}, \bibinfo {author} {\bibfnamefont {A.~F.}\
  \bibnamefont {Kockum}}, \bibinfo {author} {\bibfnamefont {S.}~\bibnamefont
  {Watabe}}, \bibinfo {author} {\bibfnamefont {F.}~\bibnamefont {Yoshihara}},\
  and\ \bibinfo {author} {\bibfnamefont {J.-S.}\ \bibnamefont {Tsai}},\ }\href
  {https://doi.org/10.48550/arXiv.2406.17999} {\bibfield  {journal} {\bibinfo
  {journal} {arXiv:2406.17999}\ } (\bibinfo {year} {2024})}\BibitemShut
  {NoStop}%
\bibitem [{\citenamefont {Puri}\ \emph {et~al.}(2020)\citenamefont {Puri},
  \citenamefont {St-Jean}, \citenamefont {Gross}, \citenamefont {Grimm},
  \citenamefont {Frattini}, \citenamefont {Iyer}, \citenamefont {Krishna},
  \citenamefont {Touzard}, \citenamefont {Jiang}, \citenamefont {Blais},
  \citenamefont {Flammia},\ and\ \citenamefont
  {Girvin}}]{KCPuri2020SciAdvBPCX}%
  \BibitemOpen
  \bibfield  {author} {\bibinfo {author} {\bibfnamefont {S.}~\bibnamefont
  {Puri}}, \bibinfo {author} {\bibfnamefont {L.}~\bibnamefont {St-Jean}},
  \bibinfo {author} {\bibfnamefont {J.~A.}\ \bibnamefont {Gross}}, \bibinfo
  {author} {\bibfnamefont {A.}~\bibnamefont {Grimm}}, \bibinfo {author}
  {\bibfnamefont {N.~E.}\ \bibnamefont {Frattini}}, \bibinfo {author}
  {\bibfnamefont {P.~S.}\ \bibnamefont {Iyer}}, \bibinfo {author}
  {\bibfnamefont {A.}~\bibnamefont {Krishna}}, \bibinfo {author} {\bibfnamefont
  {S.}~\bibnamefont {Touzard}}, \bibinfo {author} {\bibfnamefont
  {L.}~\bibnamefont {Jiang}}, \bibinfo {author} {\bibfnamefont
  {A.}~\bibnamefont {Blais}}, \bibinfo {author} {\bibfnamefont {S.~T.}\
  \bibnamefont {Flammia}},\ and\ \bibinfo {author} {\bibfnamefont {S.~M.}\
  \bibnamefont {Girvin}},\ }\href {https://doi.org/doi:10.1126/sciadv.aay5901}
  {\bibfield  {journal} {\bibinfo  {journal} {Science Advances}\ }\textbf
  {\bibinfo {volume} {6}},\ \bibinfo {pages} {eaay5901} (\bibinfo {year}
  {2020})}\BibitemShut {NoStop}%
\bibitem [{\citenamefont {Hua}\ \emph {et~al.}(2024)\citenamefont {Hua},
  \citenamefont {Xu}, \citenamefont {Wang}, \citenamefont {Ma}, \citenamefont
  {Zhou}, \citenamefont {Cai}, \citenamefont {Ai}, \citenamefont {Liu},
  \citenamefont {Li}, \citenamefont {Zou},\ and\ \citenamefont {Sun}}]{NEMS}%
  \BibitemOpen
  \bibfield  {author} {\bibinfo {author} {\bibfnamefont {Z.}~\bibnamefont
  {Hua}}, \bibinfo {author} {\bibfnamefont {Y.}~\bibnamefont {Xu}}, \bibinfo
  {author} {\bibfnamefont {W.}~\bibnamefont {Wang}}, \bibinfo {author}
  {\bibfnamefont {Y.}~\bibnamefont {Ma}}, \bibinfo {author} {\bibfnamefont
  {J.}~\bibnamefont {Zhou}}, \bibinfo {author} {\bibfnamefont {W.}~\bibnamefont
  {Cai}}, \bibinfo {author} {\bibfnamefont {H.}~\bibnamefont {Ai}}, \bibinfo
  {author} {\bibfnamefont {Y.-X.}\ \bibnamefont {Liu}}, \bibinfo {author}
  {\bibfnamefont {M.}~\bibnamefont {Li}}, \bibinfo {author} {\bibfnamefont
  {C.-L.}\ \bibnamefont {Zou}},\ and\ \bibinfo {author} {\bibfnamefont
  {L.}~\bibnamefont {Sun}},\ }\href {https://arxiv.org/abs/2410.06904}
  {\bibfield  {journal} {\bibinfo  {journal} {arXiv:2410.06904}\ } (\bibinfo
  {year} {2024})}\BibitemShut {NoStop}%
\bibitem [{\citenamefont {Frattini}\ \emph {et~al.}(2017)\citenamefont
  {Frattini}, \citenamefont {Vool}, \citenamefont {Shankar}, \citenamefont
  {Narla}, \citenamefont {Sliwa},\ and\ \citenamefont
  {Devoret}}]{SNAILFrattini2017APL3waveMixing}%
  \BibitemOpen
  \bibfield  {author} {\bibinfo {author} {\bibfnamefont {N.~E.}\ \bibnamefont
  {Frattini}}, \bibinfo {author} {\bibfnamefont {U.}~\bibnamefont {Vool}},
  \bibinfo {author} {\bibfnamefont {S.}~\bibnamefont {Shankar}}, \bibinfo
  {author} {\bibfnamefont {A.}~\bibnamefont {Narla}}, \bibinfo {author}
  {\bibfnamefont {K.~M.}\ \bibnamefont {Sliwa}},\ and\ \bibinfo {author}
  {\bibfnamefont {M.~H.}\ \bibnamefont {Devoret}},\ }\href
  {https://doi.org/10.1063/1.4984142} {\bibfield  {journal} {\bibinfo
  {journal} {Applied Physics Letters}\ }\textbf {\bibinfo {volume} {110}},\
  \bibinfo {pages} {222603} (\bibinfo {year} {2017})}\BibitemShut {NoStop}%
\bibitem [{\citenamefont {Frattini}\ \emph {et~al.}(2018)\citenamefont
  {Frattini}, \citenamefont {Sivak}, \citenamefont {Lingenfelter},
  \citenamefont {Shankar},\ and\ \citenamefont
  {Devoret}}]{SNAIL2018FrattiniPRATUPA}%
  \BibitemOpen
  \bibfield  {author} {\bibinfo {author} {\bibfnamefont {N.~E.}\ \bibnamefont
  {Frattini}}, \bibinfo {author} {\bibfnamefont {V.~V.}\ \bibnamefont {Sivak}},
  \bibinfo {author} {\bibfnamefont {A.}~\bibnamefont {Lingenfelter}}, \bibinfo
  {author} {\bibfnamefont {S.}~\bibnamefont {Shankar}},\ and\ \bibinfo {author}
  {\bibfnamefont {M.~H.}\ \bibnamefont {Devoret}},\ }\href
  {https://doi.org/10.1103/PhysRevApplied.10.054020} {\bibfield  {journal}
  {\bibinfo  {journal} {Physical Review Applied}\ }\textbf {\bibinfo {volume}
  {10}},\ \bibinfo {pages} {054020} (\bibinfo {year} {2018})}\BibitemShut
  {NoStop}%
\bibitem [{\citenamefont {Sivak}\ \emph {et~al.}(2019)\citenamefont {Sivak},
  \citenamefont {Frattini}, \citenamefont {Joshi}, \citenamefont
  {Lingenfelter}, \citenamefont {Shankar},\ and\ \citenamefont
  {Devoret}}]{SNAILSivak2019PRAKerrFree}%
  \BibitemOpen
  \bibfield  {author} {\bibinfo {author} {\bibfnamefont {V.~V.}\ \bibnamefont
  {Sivak}}, \bibinfo {author} {\bibfnamefont {N.~E.}\ \bibnamefont {Frattini}},
  \bibinfo {author} {\bibfnamefont {V.~R.}\ \bibnamefont {Joshi}}, \bibinfo
  {author} {\bibfnamefont {A.}~\bibnamefont {Lingenfelter}}, \bibinfo {author}
  {\bibfnamefont {S.}~\bibnamefont {Shankar}},\ and\ \bibinfo {author}
  {\bibfnamefont {M.~H.}\ \bibnamefont {Devoret}},\ }\href
  {https://doi.org/10.1103/PhysRevApplied.11.054060} {\bibfield  {journal}
  {\bibinfo  {journal} {Physical Review Applied}\ }\textbf {\bibinfo {volume}
  {11}},\ \bibinfo {pages} {054060} (\bibinfo {year} {2019})}\BibitemShut
  {NoStop}%
\bibitem [{\citenamefont {He}\ \emph {et~al.}(2023)\citenamefont {He},
  \citenamefont {Lu}, \citenamefont {Bao}, \citenamefont {Xue}, \citenamefont
  {Jiang}, \citenamefont {Wang}, \citenamefont {Roudsari}, \citenamefont
  {Delsing}, \citenamefont {Tsai},\ and\ \citenamefont {Lin}}]{KCHe2023NCK0}%
  \BibitemOpen
  \bibfield  {author} {\bibinfo {author} {\bibfnamefont {X.~L.}\ \bibnamefont
  {He}}, \bibinfo {author} {\bibfnamefont {Y.}~\bibnamefont {Lu}}, \bibinfo
  {author} {\bibfnamefont {D.~Q.}\ \bibnamefont {Bao}}, \bibinfo {author}
  {\bibfnamefont {H.}~\bibnamefont {Xue}}, \bibinfo {author} {\bibfnamefont
  {W.~B.}\ \bibnamefont {Jiang}}, \bibinfo {author} {\bibfnamefont
  {Z.}~\bibnamefont {Wang}}, \bibinfo {author} {\bibfnamefont {A.~F.}\
  \bibnamefont {Roudsari}}, \bibinfo {author} {\bibfnamefont {P.}~\bibnamefont
  {Delsing}}, \bibinfo {author} {\bibfnamefont {J.~S.}\ \bibnamefont {Tsai}},\
  and\ \bibinfo {author} {\bibfnamefont {Z.~R.}\ \bibnamefont {Lin}},\ }\href
  {https://doi.org/10.1038/s41467-023-42057-0} {\bibfield  {journal} {\bibinfo
  {journal} {Nature Communications}\ }\textbf {\bibinfo {volume} {14}},\
  \bibinfo {pages} {6358} (\bibinfo {year} {2023})}\BibitemShut {NoStop}%
\bibitem [{\citenamefont {Bergeal}\ \emph {et~al.}(2010)\citenamefont
  {Bergeal}, \citenamefont {Vijay}, \citenamefont {Manucharyan}, \citenamefont
  {Siddiqi}, \citenamefont {Schoelkopf}, \citenamefont {Girvin},\ and\
  \citenamefont {Devoret}}]{JRMBergeal2010NP}%
  \BibitemOpen
  \bibfield  {author} {\bibinfo {author} {\bibfnamefont {N.}~\bibnamefont
  {Bergeal}}, \bibinfo {author} {\bibfnamefont {R.}~\bibnamefont {Vijay}},
  \bibinfo {author} {\bibfnamefont {V.~E.}\ \bibnamefont {Manucharyan}},
  \bibinfo {author} {\bibfnamefont {I.}~\bibnamefont {Siddiqi}}, \bibinfo
  {author} {\bibfnamefont {R.~J.}\ \bibnamefont {Schoelkopf}}, \bibinfo
  {author} {\bibfnamefont {S.~M.}\ \bibnamefont {Girvin}},\ and\ \bibinfo
  {author} {\bibfnamefont {M.~H.}\ \bibnamefont {Devoret}},\ }\href
  {https://doi.org/10.1038/nphys1516} {\bibfield  {journal} {\bibinfo
  {journal} {Nature Physics}\ }\textbf {\bibinfo {volume} {6}},\ \bibinfo
  {pages} {296} (\bibinfo {year} {2010})}\BibitemShut {NoStop}%
\bibitem [{\citenamefont {Roch}\ \emph {et~al.}(2012)\citenamefont {Roch},
  \citenamefont {Flurin}, \citenamefont {Nguyen}, \citenamefont {Morfin},
  \citenamefont {Campagne-Ibarcq}, \citenamefont {Devoret},\ and\ \citenamefont
  {Huard}}]{JRMRoch2012PRL}%
  \BibitemOpen
  \bibfield  {author} {\bibinfo {author} {\bibfnamefont {N.}~\bibnamefont
  {Roch}}, \bibinfo {author} {\bibfnamefont {E.}~\bibnamefont {Flurin}},
  \bibinfo {author} {\bibfnamefont {F.}~\bibnamefont {Nguyen}}, \bibinfo
  {author} {\bibfnamefont {P.}~\bibnamefont {Morfin}}, \bibinfo {author}
  {\bibfnamefont {P.}~\bibnamefont {Campagne-Ibarcq}}, \bibinfo {author}
  {\bibfnamefont {M.~H.}\ \bibnamefont {Devoret}},\ and\ \bibinfo {author}
  {\bibfnamefont {B.}~\bibnamefont {Huard}},\ }\href
  {https://doi.org/10.1103/PhysRevLett.108.147701} {\bibfield  {journal}
  {\bibinfo  {journal} {Physical Review Letters}\ }\textbf {\bibinfo {volume}
  {108}},\ \bibinfo {pages} {147701} (\bibinfo {year} {2012})}\BibitemShut
  {NoStop}%
\bibitem [{SM()}]{SM}%
  \BibitemOpen
  \href@noop {} {\bibinfo {title} {See supplemental material for details on the
  system parameters, the calibration of devices, and the characterization of
  kerr-cat qubit}}\BibitemShut {NoStop}%
\bibitem [{\citenamefont {Yanagimoto}\ \emph {et~al.}(2019)\citenamefont
  {Yanagimoto}, \citenamefont {Ng}, \citenamefont {Onodera},\ and\
  \citenamefont {Mabuchi}}]{StatePreYanagimoto2019PRAFock}%
  \BibitemOpen
  \bibfield  {author} {\bibinfo {author} {\bibfnamefont {R.}~\bibnamefont
  {Yanagimoto}}, \bibinfo {author} {\bibfnamefont {E.}~\bibnamefont {Ng}},
  \bibinfo {author} {\bibfnamefont {T.}~\bibnamefont {Onodera}},\ and\ \bibinfo
  {author} {\bibfnamefont {H.}~\bibnamefont {Mabuchi}},\ }\href
  {https://doi.org/10.1103/PhysRevA.100.033822} {\bibfield  {journal} {\bibinfo
   {journal} {Physical Review A}\ }\textbf {\bibinfo {volume} {100}},\ \bibinfo
  {pages} {033822} (\bibinfo {year} {2019})}\BibitemShut {NoStop}%
\bibitem [{\citenamefont {Dirac}\ and\ \citenamefont
  {Rutherford}(1925)}]{Dirac1925PRSAdiabatic}%
  \BibitemOpen
  \bibfield  {author} {\bibinfo {author} {\bibfnamefont {P.~A.~M.}\
  \bibnamefont {Dirac}}\ and\ \bibinfo {author} {\bibfnamefont
  {E.}~\bibnamefont {Rutherford}},\ }\href
  {https://doi.org/doi:10.1098/rspa.1925.0052} {\bibfield  {journal} {\bibinfo
  {journal} {Proceedings of the Royal Society of London. Series A, Containing
  Papers of a Mathematical and Physical Character}\ }\textbf {\bibinfo {volume}
  {107}},\ \bibinfo {pages} {725} (\bibinfo {year} {1925})}\BibitemShut
  {NoStop}%
\bibitem [{\citenamefont {Sun}\ \emph {et~al.}(2014)\citenamefont {Sun},
  \citenamefont {Petrenko}, \citenamefont {Leghtas}, \citenamefont {Vlastakis},
  \citenamefont {Kirchmair}, \citenamefont {Sliwa}, \citenamefont {Narla},
  \citenamefont {Hatridge}, \citenamefont {Shankar}, \citenamefont {Blumoff},
  \citenamefont {Frunzio}, \citenamefont {Mirrahimi}, \citenamefont {Devoret},\
  and\ \citenamefont {Schoelkopf}}]{ParitySun2014Nature}%
  \BibitemOpen
  \bibfield  {author} {\bibinfo {author} {\bibfnamefont {L.}~\bibnamefont
  {Sun}}, \bibinfo {author} {\bibfnamefont {A.}~\bibnamefont {Petrenko}},
  \bibinfo {author} {\bibfnamefont {Z.}~\bibnamefont {Leghtas}}, \bibinfo
  {author} {\bibfnamefont {B.}~\bibnamefont {Vlastakis}}, \bibinfo {author}
  {\bibfnamefont {G.}~\bibnamefont {Kirchmair}}, \bibinfo {author}
  {\bibfnamefont {K.~M.}\ \bibnamefont {Sliwa}}, \bibinfo {author}
  {\bibfnamefont {A.}~\bibnamefont {Narla}}, \bibinfo {author} {\bibfnamefont
  {M.}~\bibnamefont {Hatridge}}, \bibinfo {author} {\bibfnamefont
  {S.}~\bibnamefont {Shankar}}, \bibinfo {author} {\bibfnamefont
  {J.}~\bibnamefont {Blumoff}}, \bibinfo {author} {\bibfnamefont
  {L.}~\bibnamefont {Frunzio}}, \bibinfo {author} {\bibfnamefont
  {M.}~\bibnamefont {Mirrahimi}}, \bibinfo {author} {\bibfnamefont {M.~H.}\
  \bibnamefont {Devoret}},\ and\ \bibinfo {author} {\bibfnamefont {R.~J.}\
  \bibnamefont {Schoelkopf}},\ }\href {https://doi.org/10.1038/nature13436}
  {\bibfield  {journal} {\bibinfo  {journal} {Nature}\ }\textbf {\bibinfo
  {volume} {511}},\ \bibinfo {pages} {444} (\bibinfo {year}
  {2014})}\BibitemShut {NoStop}%
\bibitem [{\citenamefont {Nielsen}\ and\ \citenamefont
  {Chuang}(2010)}]{NielsenChuang}%
  \BibitemOpen
  \bibfield  {author} {\bibinfo {author} {\bibfnamefont {M.~A.}\ \bibnamefont
  {Nielsen}}\ and\ \bibinfo {author} {\bibfnamefont {I.~L.}\ \bibnamefont
  {Chuang}},\ }\href {https://doi.org/DOI: 10.1017/CBO9780511976667} {\emph
  {\bibinfo {title} {Quantum Computation and Quantum Information: 10th
  Anniversary Edition}}}\ (\bibinfo  {publisher} {Cambridge University Press},\
  \bibinfo {address} {Cambridge},\ \bibinfo {year} {2010})\BibitemShut
  {NoStop}%
\bibitem [{\citenamefont {Landau}(1932)}]{Landau1932PZSLZtransitions}%
  \BibitemOpen
  \bibfield  {author} {\bibinfo {author} {\bibfnamefont {L.}~\bibnamefont
  {Landau}},\ }\href@noop {} {\bibfield  {journal} {\bibinfo  {journal}
  {Physikalische Zeitschrift der Sowjetunion}\ }\textbf {\bibinfo {volume}
  {2}},\ \bibinfo {pages} {46} (\bibinfo {year} {1932})}\BibitemShut {NoStop}%
\bibitem [{\citenamefont {Zener}\ and\ \citenamefont
  {Fowler}(1932)}]{Zener1932PRSLZtransitions}%
  \BibitemOpen
  \bibfield  {author} {\bibinfo {author} {\bibfnamefont {C.}~\bibnamefont
  {Zener}}\ and\ \bibinfo {author} {\bibfnamefont {R.~H.}\ \bibnamefont
  {Fowler}},\ }\href {https://doi.org/doi:10.1098/rspa.1932.0165} {\bibfield
  {journal} {\bibinfo  {journal} {Proceedings of the Royal Society of London.
  Series A, Containing Papers of a Mathematical and Physical Character}\
  }\textbf {\bibinfo {volume} {137}},\ \bibinfo {pages} {696} (\bibinfo {year}
  {1932})}\BibitemShut {NoStop}%
\bibitem [{\citenamefont {Johansson}\ \emph {et~al.}(2012)\citenamefont
  {Johansson}, \citenamefont {Nation},\ and\ \citenamefont {Nori}}]{qutip1}%
  \BibitemOpen
  \bibfield  {author} {\bibinfo {author} {\bibfnamefont {J.~R.}\ \bibnamefont
  {Johansson}}, \bibinfo {author} {\bibfnamefont {P.~D.}\ \bibnamefont
  {Nation}},\ and\ \bibinfo {author} {\bibfnamefont {F.}~\bibnamefont {Nori}},\
  }\href {https://doi.org/https://doi.org/10.1016/j.cpc.2012.02.021} {\bibfield
   {journal} {\bibinfo  {journal} {Computer Physics Communications}\ }\textbf
  {\bibinfo {volume} {183}},\ \bibinfo {pages} {1760} (\bibinfo {year}
  {2012})}\BibitemShut {NoStop}%
\bibitem [{\citenamefont {Johansson}\ \emph {et~al.}(2013)\citenamefont
  {Johansson}, \citenamefont {Nation},\ and\ \citenamefont {Nori}}]{qutip2}%
  \BibitemOpen
  \bibfield  {author} {\bibinfo {author} {\bibfnamefont {J.~R.}\ \bibnamefont
  {Johansson}}, \bibinfo {author} {\bibfnamefont {P.~D.}\ \bibnamefont
  {Nation}},\ and\ \bibinfo {author} {\bibfnamefont {F.}~\bibnamefont {Nori}},\
  }\href {https://doi.org/https://doi.org/10.1016/j.cpc.2012.11.019} {\bibfield
   {journal} {\bibinfo  {journal} {Computer Physics Communications}\ }\textbf
  {\bibinfo {volume} {184}},\ \bibinfo {pages} {1234} (\bibinfo {year}
  {2013})}\BibitemShut {NoStop}%
\bibitem [{\citenamefont {Gautier}\ \emph {et~al.}(2022)\citenamefont
  {Gautier}, \citenamefont {Sarlette},\ and\ \citenamefont
  {Mirrahimi}}]{KDCGautier2022PRXQ}%
  \BibitemOpen
  \bibfield  {author} {\bibinfo {author} {\bibfnamefont {R.}~\bibnamefont
  {Gautier}}, \bibinfo {author} {\bibfnamefont {A.}~\bibnamefont {Sarlette}},\
  and\ \bibinfo {author} {\bibfnamefont {M.}~\bibnamefont {Mirrahimi}},\ }\href
  {https://doi.org/10.1103/PRXQuantum.3.020339} {\bibfield  {journal} {\bibinfo
   {journal} {PRX Quantum}\ }\textbf {\bibinfo {volume} {3}},\ \bibinfo {pages}
  {020339} (\bibinfo {year} {2022})}\BibitemShut {NoStop}%
\end{thebibliography}

%

\end{document}


\title{Supplementary Materials to: ``Dynamic Compensation for Pump-Induced Frequency Shift in Kerr-Cat Qubit Initialization"}

\author{Yifang Xu}
\thanks{These authors contributed equally to this work.}
\author{Ziyue Hua}
\thanks{These authors contributed equally to this work.}
\author{Weiting Wang}
\email{wangwt2020@tsinghua.edu.cn}
\affiliation{Center for Quantum Information, Institute for Interdisciplinary, Information Sciences, Tsinghua University, Beijing 100084, China}

\author{Yuwei Ma}
\affiliation{Center for Quantum Information, Institute for Interdisciplinary, Information Sciences, Tsinghua University, Beijing 100084, China}

\author{Ming Li}
\affiliation{CAS Key Laboratory of Quantum Information, University of Science and Technology of China, Hefei 230026, China}

\author{Jiajun Chen}
\affiliation{Center for Quantum Information, Institute for Interdisciplinary, Information Sciences, Tsinghua University, Beijing 100084, China}

\author{Jie~Zhou}
\affiliation{Center for Quantum Information, Institute for Interdisciplinary, Information Sciences, Tsinghua University, Beijing 100084, China}

\author{Xiaoxuan~Pan}
\affiliation{Center for Quantum Information, Institute for Interdisciplinary, Information Sciences, Tsinghua University, Beijing 100084, China}

\author{Lintao~Xiao}
\affiliation{Center for Quantum Information, Institute for Interdisciplinary, Information Sciences, Tsinghua University, Beijing 100084, China}

\author{Hongwei~Huang}
\affiliation{Center for Quantum Information, Institute for Interdisciplinary, Information Sciences, Tsinghua University, Beijing 100084, China}

\author{Weizhou Cai}
\affiliation{CAS Key Laboratory of Quantum Information, University of Science and Technology of China, Hefei 230026, China}

\author{Hao Ai}
\affiliation{Institute of Microelectronics, Tsinghua University, Beijing 100084, China}


\author{Yu-xi Liu}
\affiliation{Institute of Microelectronics, Tsinghua University, Beijing 100084, China}

\author{Chang-Ling Zou}
\email{clzou321@ustc.edu.cn}
\affiliation{CAS Key Laboratory of Quantum Information, University of Science and Technology of China, Hefei 230026, China}
\affiliation{Hefei National Laboratory, Hefei 230088, China}

\author{Luyan Sun}
\email{luyansun@tsinghua.edu.cn}
\affiliation{Center for Quantum Information, Institute for Interdisciplinary, Information Sciences, Tsinghua University, Beijing 100084, China}
\affiliation{Hefei National Laboratory, Hefei 230088, China}


\maketitle

\tableofcontents

\newpage

\renewcommand{\thefigure}{S\arabic{figure}}
\setcounter{figure}{0}
\renewcommand{\thetable}{S\arabic{table}}
\setcounter{table}{0}

\section{The Nonlinearity-Engineered Multi-loop SQUID}\label{NEMS Hamiltonian}

The \textit{Nonlinearity-Engineered Multi-loop SQUID} (NEMS)~\cite{NEMS} is a novel superconducting qubit that enables the engineering of the second-order nonlinearity for potential quantum devices. In this section, we briefly explore the Hamiltonian, describe the NEMS qubit~\cite{NEMS}, and analyze the pump-induced frequency shift (PIFS). There are two main contributions to the PIFS. One contribution is known as the ``AC Stark shift", which comes from the charge drive~\cite{KCGrimm2020Nature}, while the other is referred to as the ``Zeeman-like shift", which results from the flux pump~\cite{FPLu2023NCBeamSplitting,KCIyama2024NCWigner}. The former contribution has already been studied in Ref.~\cite{KCGrimm2020Nature} and it is revealed that its value is proportional to the square of the pump strength. Therefore, we focus on the latter one in the following, and we prove that its value is also proportional to the square of the pump strength. As a consequence, we do not need to distinguish these two mechanisms in our experiments.

\begin{figure}[htbp]
	\centering
	\includegraphics{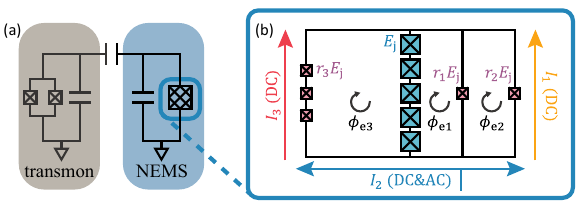}
	\caption{(a) Circuit schematic of the device: a NEMS is coupled with a transmon. The gray part is the transmon, serving as an ancilla. The blue part is the NEMS. 
	(b) Circuit schematic of the NEMS. The Josephson energies for large junctions are $E_\mathrm{j}$, while for small junctions they are $r_1E_\mathrm{j}$, $r_2E_\mathrm{j}$, and $r_3E_\mathrm{j}$, respectively. We design $r_1=r_2=r_3=r$ for the device studied in this work. Three current lines can completely control external fluxes in the three loops. The AC flux ratios of the three loops induced by $I_2$ is carefully designed as 1:-1:3, where the negative sign refers to the different direction of the magnetic field.}
	\label{fig:circuit}  
\end{figure}

Figure~\ref{fig:circuit}(a) schematically illustrates the circuit of the device investigated in our experiments, where the NEMS is coupled to a transmon qubit. Since the transmon qubit is an ancillary qubit, we only concentrate on the NEMS in this section. The detailed circuit of the NEMS is shown in Fig.~\ref{fig:circuit}(b), featuring three loops interconnected by multiple Josephson junctions. The corresponding Hamiltonian that describes the NEMS reads
\begin{equation}\label{eq:NEMSHamiltonian}
	\hat{H}_\mathrm{NEMS} = \hat{U}_\mathrm{NEMS} + 4E_\mathrm{c}\hat{n}^2,
\end{equation}
with the potential energy determined by the fluxes ($\phi_\mathrm{e1,e2,e3}$) threading the loops as
\begin{equation}\label{eq:NEMSPotentialEnergy}
    \hat{U}_\mathrm{NEMS}/E_\mathrm{j} = -r_1\cos(\hat{\varphi}+\phi_\mathrm{e1})
    					 -r_2\cos(\hat{\varphi}+\phi_\mathrm{e1}+\phi_\mathrm{e2})
    					 -n_rr_3\cos(\frac{\hat{\varphi}-\phi_\mathrm{e3}}{n_r})
    					 -n_\mathrm{L}\cos(\frac{\hat{\varphi}}{n_\mathrm{L}}).
\end{equation}
Here, $\hat{n}$ and $\hat{\varphi}$ are the charge and flux operators, respectively, satisfying $[\hat{n},\hat{\varphi}]=1$. $E_\mathrm{c}$ is the charging energy, $E_\mathrm{j}$ is the Josephson energy of the large junction, while $r_1$, $r_2$, and $r_3$ specify the ratios of the Josephson energies for the small junctions. $n_r$ and $n_\mathrm{L}$ denote the number of junctions in the loops that are determined by the specific device geometry, and we have $n_r=3$ and $n_\mathrm{L}=5$ in this work. The fluxes ($\phi_\mathrm{e1,e2,e3}$) are controlled by three current lines, and the ratios of these three fluxes can be designed based on the loop size and the position of the flux pump line.

We first bias the device around the working point  $(\phi_\mathrm{e1},\phi_\mathrm{e2},\phi_\mathrm{e3}) = (0,\pi,0)$ to cancel the first-order AC term, which is required for realizing Kerr-cat CNOT gate~\cite{KCPuri2020SciAdvBPCX,NEMS}. To generate the second-order AC term in the NEMS, we introduce a small shift $\Delta\phi$ to this working point and apply a flux pump such that 
\begin{eqnarray}\label{eq:workingpoint}
    \phi_\mathrm{e1} &=& 0 - \Delta\phi + \varepsilon,\label{eq:phie1}\\
     \phi_\mathrm{e2}& =& \pi + \Delta\phi - \varepsilon,\label{eq:phie2}\\
      \phi_\mathrm{e3} &= &0 + n_r\varepsilon.\label{eq:phie3}
\end{eqnarray}
Here, $\varepsilon = \epsilon_\mathrm{p}\mathrm{cos}(\omega_\mathrm{p} t)$ is due to the flux pump, and we set $\Delta\phi=0.08\pi$ in the subsequent experiments. Substituting Eqs.~(\ref{eq:phie1}-\ref{eq:phie3}) into Eq.~(\ref{eq:NEMSPotentialEnergy}), we obtain
\begin{equation}
    \hat{U}_\mathrm{NEMS}/E_\mathrm{j} = -r_1\cos(\hat{\varphi} - \Delta\phi + \varepsilon)
    					 -r_2\cos(\hat{\varphi})
    					 -n_rr_3\cos(\frac{\hat{\varphi}}{n_r} - \varepsilon)
    					 -n_\mathrm{L}\cos(\frac{\hat{\varphi}}{n_\mathrm{L}}),
\label{eq:NEMS_PE_working}
\end{equation}
which can be divided into the static and driven-induced parts as
\begin{equation}
    \hat{U}_\mathrm{NEMS} = \hat{U}^\mathrm{Static}_\mathrm{NEMS} + \hat{U}^\mathrm{Driven}_\mathrm{NEMS},
\end{equation}
with
\begin{equation}
    \hat{U}^\mathrm{Static}_\mathrm{NEMS}/E_\mathrm{j} =
	-r_1\cos(\hat{\varphi} - \Delta\phi )
	-r_2\cos(\hat{\varphi})
	-n_rr_3\cos(\frac{\hat{\varphi}}{n_r})
	-n_\mathrm{L}\cos(\frac{\hat{\varphi}}{n_\mathrm{L}}),
\label{eq:NEMS_PE_Static}
\end{equation}
and 
\begin{equation}
    \hat{U}^\mathrm{Driven}_\mathrm{NEMS}/E_\mathrm{j} =
    	\bigg[r_1\sin(\hat{\varphi} - \Delta\phi )
    	-n_rr_3\sin(\frac{\hat{\varphi}}{n_r})\bigg]\sin\varepsilon
    	+\bigg[-r_1\cos(\hat{\varphi} - \Delta\phi )
    	-n_rr_3\cos(\frac{\hat{\varphi}}{n_r})\bigg](\cos\varepsilon-1).
\label{eq:NEMS_PE_Driven}
\end{equation}

For a small pump amplitude $\epsilon_\mathrm{p}$, the driven term could be approximated by the Jacob-Anger expansion to the second-order as 
$\sin(\epsilon_\mathrm{p} \cos(\omega_\mathrm{p} t)) \simeq 2J_1(\epsilon_\mathrm{p})\cos(\omega_\mathrm{p} t)$,
$\cos(\epsilon_\mathrm{p} \cos(\omega_\mathrm{p} t))-1 \simeq [J_0(\epsilon_\mathrm{p})-1] - 2J_2(\epsilon_\mathrm{p})\cos(2\omega_\mathrm{p} t)$. 
Consequently, we can rewrite the potential energy as
\begin{equation}\label{eq:U_D012}
    \hat{U}^\mathrm{Driven}_\mathrm{NEMS} = \hat{U}^\mathrm{Driven,(0)}_\mathrm{NEMS} + \hat{U}^\mathrm{Driven,(1)}_\mathrm{NEMS} + \hat{U}^\mathrm{Driven,(2)}_\mathrm{NEMS} + ...,
\end{equation}
with the zero-th order term
\begin{equation}\label{eq:UD0}
    \hat{U}^\mathrm{Driven,(0)}_\mathrm{NEMS}/E_\mathrm{j} =
	\bigg[-r_1\cos(\hat{\varphi} - \Delta\phi )
   	-n_rr_3\cos(\frac{\hat{\varphi}}{n_r})\bigg]\bigg[-(\frac{\epsilon_\mathrm{p}}{2})^2+O(\epsilon_\mathrm{p}^4)\bigg],
\end{equation}
the first order term
\begin{equation}\label{eq:UD1}
    \hat{U}^\mathrm{Driven,(1)}_\mathrm{NEMS}/E_\mathrm{j} =
	\bigg[r_1\sin(\hat{\varphi} - \Delta\phi )
   	-n_rr_3\sin(\frac{\hat{\varphi}}{n_r})\bigg]\bigg[\epsilon_\mathrm{p}-(\frac{\epsilon_\mathrm{p}}{2})^3+O(\epsilon_\mathrm{p}^5)\bigg]\cos(\omega_\mathrm{p} t),
\end{equation}
and the second order term
\begin{equation}\label{eq:UD2}
    \hat{U}^\mathrm{Driven,(2)}_\mathrm{NEMS}/E_\mathrm{j} =
	\bigg[-r_1\cos(\hat{\varphi} - \Delta\phi )
   	-n_rr_3\cos(\frac{\hat{\varphi}}{n_r})\bigg]\bigg[-(\frac{\epsilon_\mathrm{p}}{2})^2+O(\epsilon_\mathrm{p}^4)\bigg]\cos(2\omega_\mathrm{p} t).
\end{equation}

The static shift in the potential energy due to the flux pump, i.e., Eq.~(\ref{eq:UD0}), contributes to the so-called ``Zeeman-like shift". It modifies both the shape and the minimum position of the potential well. 

Then, the total static potential energy by summing up Eq.~(\ref{eq:NEMS_PE_Static}) and Eq.~(\ref{eq:UD0}) reads
\begin{equation}\label{eq:U_S_FP}
    \hat{U}^\mathrm{Static,FP}_\mathrm{NEMS}/E_\mathrm{j} =
	    -r_1\cos(\hat{\varphi} - \Delta\phi )\bigg[1-(\frac{\epsilon_\mathrm{p}}{2})^2\bigg]
	   	-r_2\cos(\hat{\varphi})
	   	-n_rr_3\cos(\frac{\hat{\varphi}}{n_r})\bigg[1-(\frac{\epsilon_\mathrm{p}}{2})^2\bigg]
	   	-n_\mathrm{L}\cos(\frac{\hat{\varphi}}{n_\mathrm{L}}).
\end{equation}

The standard procedure to calculate the AC Zeeman-like shift is as follows: 1. Determine the qubit frequencies from Eq.~(\ref{eq:U_S_FP}) (with flux pump) and Eq.~(\ref{eq:NEMS_PE_Static}) (without flux pump). 2. Derive the difference of the two frequencies, which equals to the AC Zeeman-like shift. In this analysis, we use the harmonic oscillator approximation (HOA). We first solve the equations to get the minimum, and then extend to the 2nd order of $\hat{\varphi}$. The second-order approximation yields a description of an inductor. Substituting the approximate potential energy back to Eq.~(\ref{eq:NEMSHamiltonian}) leads to an extraction of the NEMS frequency. Note that the HOA condition is not always satisfied for all working points (see Ref.~\cite{NEMS}).

\begin{figure}[htbp]
	\centering
	\includegraphics{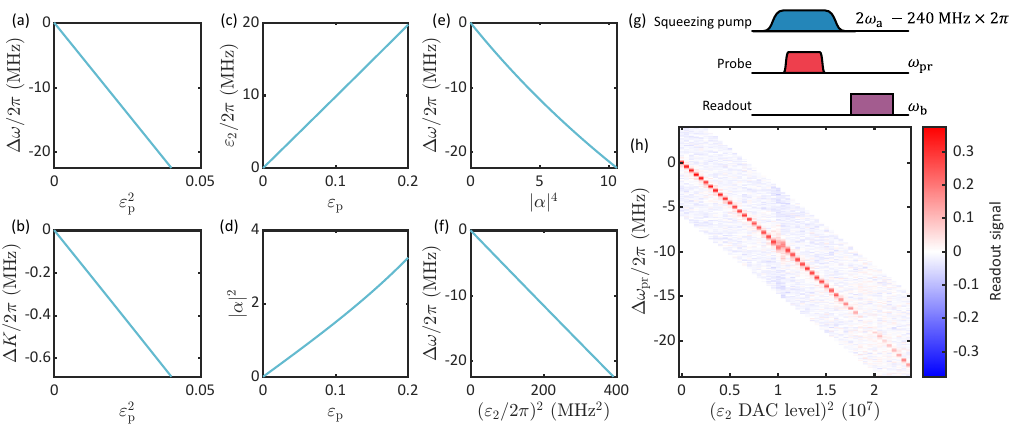}
	\caption{Numerical analysis of the AC Zeeman-like shift and other parameters with the flux pump. 
    (a-d) The dependence of the frequency shift ($\Delta\omega$, i.e., the AC Zeeman-like shift), the Kerr shift ($\Delta K$), the two-photon squeezing drive strength ($\varepsilon_2$), and the average photon number ($|\alpha|^2$, or the cat size) on the flux pump strength $\epsilon_\mathrm{p}$.
	(e, f) The dependence of the AC Zeeman-like shift on $|\alpha|^4$ and $|\varepsilon_2|^2$ is approximately linear, which is similar to the behavior in the charge pump situation~\cite{KCGrimm2020Nature}.
	(g) Experimental sequence to verify the relationship of the PIFS and the squeezing pump strength. The squeezing pump frequency is far detuned from twice the NEMS frequency to avoid entering the Kerr-cat space. By sweeping the probe frequency, we measure the NEMS frequency while applying the flux pump.
	(h) Experimental results of the PIFS. 
	}
	\label{fig:ZSsimu}  
\end{figure}

However, Eq.~(\ref{eq:NEMS_PE_Static}) and Eq.~(\ref{eq:U_S_FP}) are transcendental equations, making it challenging to obtain analytical expressions for the potential minimum and the corresponding qubit frequency. Instead, we numerically calculate the Zeeman-like shift and other involved parameters as a function of the flux pump amplitude (Fig.~\ref{fig:ZSsimu}), using the device parameters calibrated from experiments (Table~\ref{tab:allparameters}). 
The simulations reveal several features: (1) For small $\epsilon_\mathrm{p}$, the frequency shift is nearly proportional to $\epsilon_\mathrm{p}^2$ [Fig.~\ref{fig:ZSsimu}(a)]. (2) The two-photon drive strength $\varepsilon_2$ [Fig.~\ref{fig:ZSsimu}(c)] and the average photon number $|\alpha|^2$ [Fig.~\ref{fig:ZSsimu}(d)] scale linearly with $\epsilon_\mathrm{p}$. Combining these two features, we find that the AC Zeeman-like shift is nearly proportional to $|\alpha|^4$ [Fig.~\ref{fig:ZSsimu}(e)] or $|\varepsilon_2|^2$ [Fig.~\ref{fig:ZSsimu}(f)], which resembles the behavior of the AC Stark shift in the case of a charge pump drive.

To verify this relationship experimentally, we design a simple protocol, as shown in Fig.~\ref{fig:ZSsimu}(g). 
Since the AC Zeeman-like shift is independent of the pump frequency [i.e., Eq.~(\ref{eq:UD0}) does not contain $\omega_\mathrm{p}$], we set the pump relatively far detuned from $2\omega_\mathrm{a}$ to avoid entering the Kerr-cat regime, where $\omega_\mathrm{a}$ is the frequency of the NEMS when no pump is applied.
We sweep the probe tone across the NEMS frequency, and measure the spectrum, as shown in Fig.~\ref{fig:ZSsimu}(h). The measured PIFS scales quadratically with $|\varepsilon_2|$, confirming our numerical simulations in Fig.~\ref{fig:ZSsimu}(f). Deviations from a perfect parabola may be attributed to coupling with unwanted two-level systems in the device.
 
Note that due to the change of the entire potential well, the Kerr nonlinearity also shifts slightly with the pump [Fig.~\ref{fig:ZSsimu}(b)]. The small shift of Kerr nonlinearity seems to violate the linear relationship between $\varepsilon_2$ and $|\alpha|^2$, since $\varepsilon_2 = K|\alpha|^2$. 
However, for the weak flux pump strengths used in our experiments,  this effect is negligible [Fig.~\ref{fig:ZSsimu}(b)]. This also implies that a large shift of Kerr nonlinearity can be achieved by carefully designing the junction parameters of the NEMS, providing a potential route to realize a cat-qubit with an enhanced size $|\alpha|^2$ with a weaker pump.

Based on the above simulation and derivations, we can express the PIFS as
\begin{equation}\label{eq:zs}
    \Delta_\mathrm{PI} \approx \gamma\varepsilon_2^2,
\end{equation}
where $\gamma$ is the coefficient for the PIFS. Here, We treat $\varepsilon_2$ as a real number by  carefully choosing the phase of the squeezing pump~\cite{KCGrimm2020Nature}.

Pumping at the squeezing frequency $\omega_\mathrm{s}\approx2\omega_\mathrm{a}$ with an initial phase $\varphi_\mathrm{s}$, taking HOA for Eq.~(\ref{eq:NEMSHamiltonian}) and Eqs.~(\ref{eq:U_D012}-\ref{eq:U_S_FP}), and keeping terms up to the first order of $\epsilon_\mathrm{p}$ for the driven part, we get a more general expression of the Hamiltonian as ($\hbar = 1$)
\begin{equation}\label{eq:Hg2AC}
    \hat{H}_\mathrm{NEMS} = \omega_\mathrm{a}\hat{a}^\dagger\hat{a} 
    + g_4(\hat{a}^\dagger+\hat{a})^4
    + g_2^\mathrm{AC}(\hat{a}^\dagger+\hat{a})^2\epsilon_\mathrm{p}
    \cos(\omega_\mathrm{s}t + \varphi_\mathrm{s})
    - \Delta_\mathrm{PI}\hat{a}^\dagger\hat{a}.   
\end{equation}
Here, $\hat{a}^\dagger(\hat{a})$ is the harmonic oscillator creation (annihilation) operator, $g_4$ and $g_2^\mathrm{AC}$ are the higher-order nonlinearity coefficients.

To simplify the Hamiltonian, we move to a rotating frame at half the pump frequency $\omega_\mathrm{s}/2$ and have $\hat{a}\rightarrow\hat{a}\exp(-\mathrm{i}\omega_\mathrm{s}t/2\hat{a}^\dagger\hat{a})$. Considering $\cos(\omega_\mathrm{s}t + \varphi_\mathrm{s}) = [\exp(-\mathrm{i}\omega_\mathrm{s}t - \mathrm{i}\varphi_\mathrm{s})+\exp(+\mathrm{i}\omega_\mathrm{s}t + \mathrm{i}\varphi_\mathrm{s})]/2$, taking the rotating wave approximation, and dropping the fast rotating terms, we eventually arrive at the driven Kerr-cat Hamiltonian
\begin{equation}\label{eq:Hdas}
    \hat{H} = 
    - K\hat{a}^\dagger\hat{a}^\dagger\hat{a}\hat{a} 
    + \varepsilon_2\hat{a}^{\dagger2} + \varepsilon_2^{*}\hat{a}^2
    + \Delta_\mathrm{as}\hat{a}^\dagger\hat{a}
    - \Delta_\mathrm{PI}\hat{a}^\dagger\hat{a},
\end{equation}
where $\Delta_\mathrm{as} = \omega_\mathrm{a} - \omega_\mathrm{s}/2$, $K=-6g_4$, and $\varepsilon_2 = g_2^\mathrm{AC}\epsilon_\mathrm{p}\exp(\mathrm{i}\varphi_\mathrm{s})/2$. By denoting $\Delta(t) = \Delta_\mathrm{as}-\Delta_\mathrm{PI}$ and letting $\varphi_\mathrm{s}=0$ and $\varepsilon_2 = \varepsilon_2(t)$, this becomes Eq.~(1) in the main text.

\section{Experimental Setup}\label{Experimental Setup}
\subsection{Full Device Layout}\label{Full Device Layout}

\begin{figure}[htbp]
	\centering
	\includegraphics{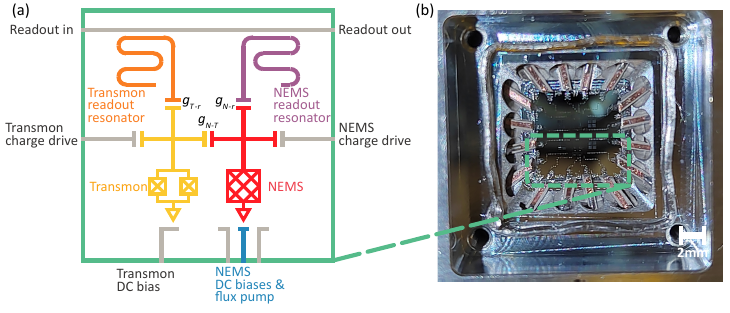}
	\caption{(a) \textbf{Full device layout.} The device contains a NEMS and a transmon, each with a readout resonator that shares the same readout transmission line, and all components are capacitively coupled. 
    (b) Photo of the wire-bonding and aluminium box with 16 ports. The green rectangle refers to (a) and the upper part is a symmetric duplication.}
	\label{fig:layout}  
\end{figure}

The full device layout is shown in Fig.~\ref{fig:layout}(a). The device consists of a transmon qubit and a NEMS, which can be controlled and read out individually. Shunted capacitors and coplanar waveguide are made of a 100~nm aluminium film on a sapphire substrate of 430~$\mathrm{\mu m}$ thick. The large Josephson junctions are fabricated with fish-bone technique \cite{FabSmith2019Thesis,KCGrimm2020Nature} and the small Josephson junctions are fabricated with the standard double-angle evaporations. 
The device is wire-bonded, packaged in an aluminium box [see Fig.~\ref{fig:layout}(b)], and put into a dilution refrigerator on the base plate at about 10~mK.

The ratios of the AC fluxes induced by $I_2$ in each loop of the NEMS are numerically simulated. Flux pump terms in Eqs.~(\ref{eq:phie1}-\ref{eq:phie3}) are highly dependent on these ratios. 
Additionally, the ratios of the DC fluxes or the \textit{$\phi$-matrix} that connects three DC currents and three DC fluxes, can be simulated following Ref.~\cite{CurrentVia2013JAP}.

\subsection{Wiring Diagram}\label{Wiring Diagram}

\begin{figure}[htbp]
	\centering
	\includegraphics{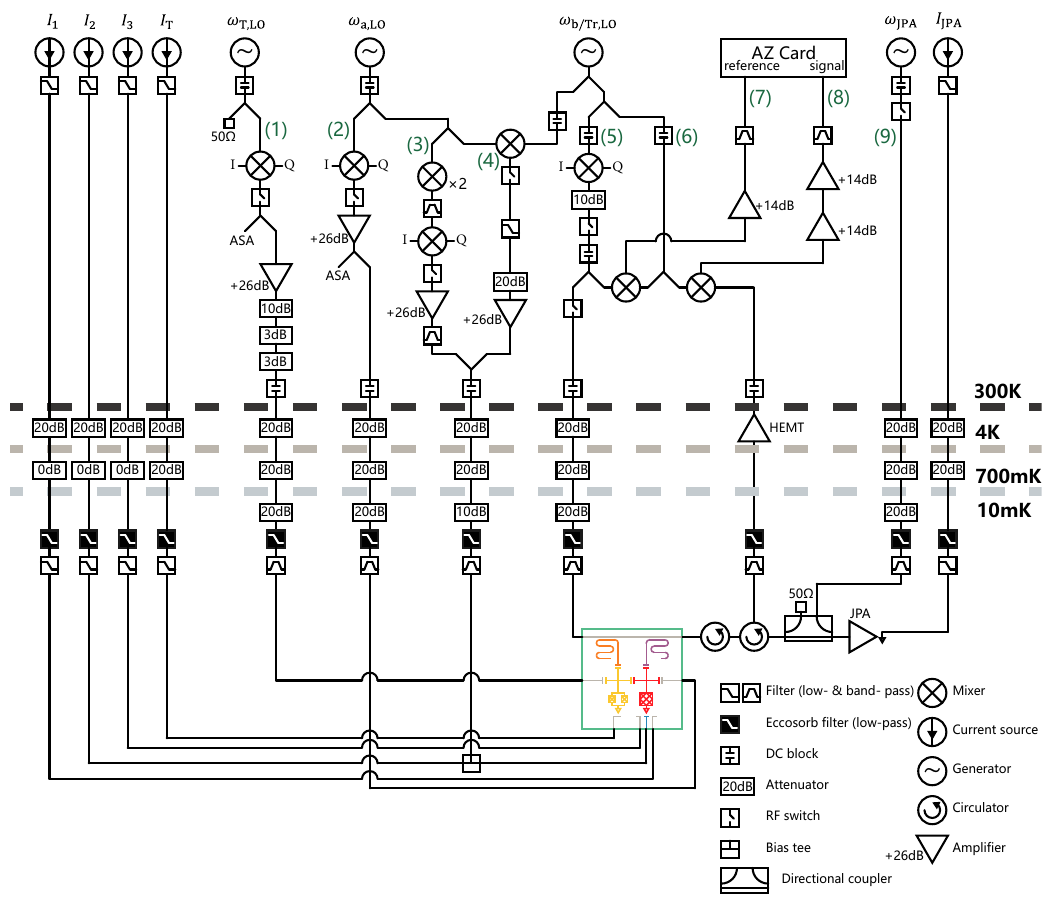}
	\caption{\textbf{Wiring diagram.} A legend for the main elements is given in the lower right corner. 
    Attenuation or gain is indicated in dB where relevant.
    The temperatures for different plates of the refrigerator are indicated on the right.
    Outside the refrigerator, wires are numbered and described in the text.
    Inside the refrigerator, a thumbnail sketch of the sample [Fig.~\ref{fig:layout}(a)] is illustrated and it is worth noting that the sample is inside a magnetic shield, which is not shown.
    }
	\label{fig:wire}  
\end{figure}

The wiring of the experiment is shown in Fig.~\ref{fig:wire}. All the lines from room temperature to the base plane have been carefully attenuated, filtered, and heat-sinked. $I_{1,2,3}$ are the current drive lines for the NEMS, $I_\mathrm{T}$ is the current drive line for the transmon, and $I_\mathrm{JPA}$ is for the Josephson parametric amplifier (JPA). 

To lock the phase, all microwave signals for NEMS are generated with two generators, $\omega_\mathrm{a,LO}$ for the qubit and $\omega_\mathrm{b,LO}$ for the readout, respectively. The microwave lines,
from left to right in Fig.~\ref{fig:wire}, are: 
(1) Charge drive for the transmon. 
(2) Charge drive for the NEMS.
(3) Double frequency flux pump (squeezing pump) for the NEMS. Notice that this line is combined with the current line $I_2$ via a bias-T at low temperature on the base plate in the refrigerator.
(4) Cat readout. This shares the flux pump port and is combined with (3) outside the refrigerator. In addition, frequencies of the microwave generators are carefully designed. The sideband for the readout is two times of that for the qubit, which makes the sideband for the cat readout zero. Hence no IQ mixer is needed in this line. Envelope of this drive is controlled by an RF switch.
(5) Dispersive readout. Both NEMS and the transmon share the same readout line. A broad band JPA is used to amplify the readout signals. Notice that there are two RF switches in this line. 
When we carry out the cat readout, the first RF switch is on while the second one is off.
(6) Local microwave signal for readout demodulation. 
(7) Readout reference.
(8) Readout signal.
(9) JPA pump.

\section{NEMS Spectrum and Parameter Calibration}\label{NEMS Spectrum and Parameters Calibration}
\subsection{Spectrum}\label{NEMS Spectrum}

\begin{figure}[htbp]
	\centering
	\includegraphics{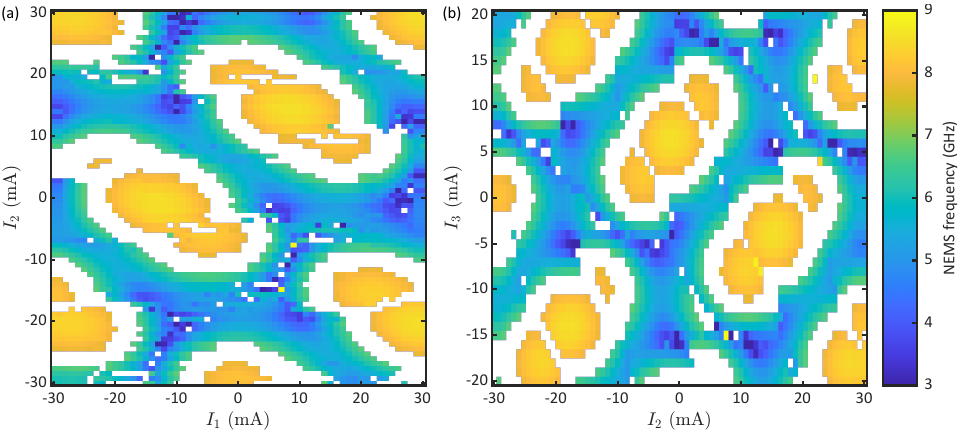}
	\caption{\textbf{NEMS Spectrum.} (a) Cross section of $I_1$ and $I_2$. (b) Cross section of $I_2$ and $I_3$.
	The NEMS frequency data presented in the figures are derived from measured readout frequencies using the Bogoliubov approach~\cite{CQEDBlais2021RMP}, along with calibrated parameters obtained through fitting. 
	The white points indicate locations where the readout signal vanishes. 
    Some of these vanishing points, primarily at the boundary of the yellow oval, indicate that the NEMS frequency passes through the readout frequency, leading to an energy level hybridization.
	Other vanishing points, mainly along the lines across the yellow oval, correspond to $\phi_\mathrm{e3}=(2k+1)\pi$, where $k$ is an integer~\cite{NEMS}. 
	}
	\label{fig:spectrum}  
\end{figure}

The NEMS spectrum is discussed in Ref.~\cite{NEMS} in terms of flux. Each flux has a period of 2$\pi$. However, in experiment our control parameters are currents. The currents  
$\boldsymbol{I}=(I_1,I_2,I_3)^\text{T} $ and external fluxes
$\boldsymbol{\phi_e}=(\phi_\mathrm{e1},\phi_\mathrm{e2},\phi_\mathrm{e3})^\text{T}$ can be linked by a $3\times3$ ``$\phi$-matrix" $M$:
\begin{equation}\label{eq:phimatrixrelationship}
    \boldsymbol{\phi_\mathrm{e}} = M (\boldsymbol{I} - I_0).
\end{equation}
Here $I_0$ is an experimental offset, which may change after each warm-up and cool-down cycle of the refrigerator.
This matrix can be simulated with the method mentioned in Ref.~\cite{CurrentVia2013JAP}, assuming the device is made of 2D superconducting thin films. 
In experiment, we sweep the three current lines to obtain the cross sections of $I_1$ and $I_2$, as shown in Fig.~\ref{fig:spectrum}(a), and $I_2$ and $I_3$, as shown in Fig.~\ref{fig:spectrum}(b).
By comparing the periods of the currents with the periods of the fluxes, we can calculate the $\phi$-matrix.

The final result of the $\phi$-matrix is 
\begin{equation}\label{eq:phimatrix}
	M = \pi
    \left[\begin{array}{c c c}
    -0.031  &  0.041  &  0.065 \\
  	-0.037 &  -0.083  &  0.070 \\
   	-0.040  &  0.200  &  0.135 
    \end{array}\right]
    (\text{mA}^{-1}).
\end{equation}

For biasing at a special point $(\phi_\mathrm{e1},\phi_\mathrm{e2},\phi_\mathrm{e3}) = (0,\pi,0)$, we adjust the currents experimentally based on the theoretical calculations from Eq.~(\ref{eq:phimatrix}) to locate the saddle point of the spectrum. 
Furthermore, we determine the additional adjustments that are needed to reach the working point as specified in Eqs.~(\ref{eq:phie1}-\ref{eq:phie3}).


\subsection{Parameter calibration}
\label{NEMS Parameters}

Procedure to calibrate the basic parameters of the device is as follows:

1. Calibrate parameters for the transmon. The ancillary transmon has a frequency of $\omega_\mathrm{T}/2\pi = 3.8\,\mathrm{GHz}$, with the amplitude damping and transverse relaxation times being $T_\mathrm{1T} = 17\,\mathrm{\mu s}$ and $T_\mathrm{2T} = 1.5\,\mathrm{\mu s}$, respectively. Since the Kerr is given by $K_\mathrm{T} = E_\mathrm{c}/2$, we get $E_\mathrm{c}$ for the transmon from a standard spectroscopy measurement. In our device, the shunted capacitors for the transmon and the NEMS are identical in size and fabrication techniques. Thus, we can also determine $E_\mathrm{c}$ for the NEMS.

2. Sweep the currents to obtain several groups of qubit and readout frequencies. Based on the Bogoliubov approach~\cite{CQEDBlais2021RMP}, we determine the coupling strength between the qubit and the readout resonator.

3. Sweep the full spectrum and identify the highest frequency point and two types of saddle points (called ``NEMS working point" and ``ATS-like point", respectively). 
According to Eq.~\ref{eq:NEMSHamiltonian} and Eq.~\ref{eq:NEMSPotentialEnergy}, the highest frequency point appears at $(\phi_\mathrm{e1},\phi_\mathrm{e2},\phi_\mathrm{e3}) = (0,0,0)$. 
``NEMS working point" refers to $(\phi_\mathrm{e1},\phi_\mathrm{e2},\phi_\mathrm{e3}) = (0,\pi,0)$, where the first order AC term is canceled.
``ATS-like point" refers to $(\phi_\mathrm{e1},\phi_\mathrm{e2},\phi_\mathrm{e3}) = (\pi,\pi,0)$, which is also a saddle point while the first order AC term is enhanced.
This feature is quite similar to that of an asymmetrically threaded SQUID (ATS)~\cite{DCLescanne2020NP}, which is why we refer to it as an ``ATS-like point".
These frequencies are given by
\begin{equation}\label{omega_highest}
    \omega_{(0,0,0)} = 
    \sqrt{ 8E_\mathrm{j}E_\mathrm{c} (r_1 + r_2 + \frac{r_3}{n_r} + \frac{1}{n_\mathrm{L}}) } 
    - \frac{ r_1 + r_2 + \frac{r_3}{n_r^3} + \frac{1}{n_\mathrm{L}^3} }{r_1 + r_2 + \frac{r_3}{n_r} + \frac{1}{n_\mathrm{L}}}E_\mathrm{c},
\end{equation}
\begin{equation}\label{omega_NEMS}
   \omega_{(0,\pi,0)} = 
       \sqrt{ 8E_\mathrm{j}E_\mathrm{c} (r_1 - r_2 + \frac{r_3}{n_r} + \frac{1}{n_\mathrm{L}}) } 
       - \frac{ r_1 - r_2 + \frac{r_3}{n_r^3} + \frac{1}{n_\mathrm{L}^3} }{r_1 - r_2 + \frac{r_3}{n_r} + \frac{1}{n_\mathrm{L}}}E_\mathrm{c},
\end{equation}
\begin{equation}\label{omega_ATS}
   \omega_{(\pi,\pi,0)} = 
       \sqrt{ 8E_\mathrm{j}E_\mathrm{c} (-r_1 - r_2 + \frac{r_3}{n_r} + \frac{1}{n_\mathrm{L}}) } 
       - \frac{ -r_1 - r_2 + \frac{r_3}{n_r^3} + \frac{1}{n_\mathrm{L}^3} }{-r_1 - r_2 + \frac{r_3}{n_r} + \frac{1}{n_\mathrm{L}}}E_\mathrm{c}.
\end{equation}
Although we design $r = r_1 = r_2 = r_3$, fluctuations in the fabrication may cause slight differences in the small junctions. 
Taking a reasonable assumption that $r = r_3 = (r_1+r_2)/2$, we can get $E_\mathrm{j}$, $r_1$, and $r_2$ based on the above three equations. 


4. Find the frequency and the Kerr for the working point shown in Eqs.~(\ref{eq:phie1}-\ref{eq:phie3}). 
Note that in this step the gates for NEMS must be weak and slow, otherwise the two-level system approximation will be invalid, as shown in Ref.~\cite{KCIyama2024NCWigner}.  

All the calibrated parameters are listed in Table \ref{tab:allparameters}.

\begin{table}
\begin{center}

\begin{tabular}{l|c|c|l} \hline \hline
\textbf{Parameters} & \textbf{Symbol} & \textbf{Value} & \textbf{Method}\\ \hline \hline
Charging energy & $E_\mathrm{c}/2\pi$ & 226~MHz & Transmon Kerr\\ \hline
Josephson energy (NEMS large junction) & $E_\mathrm{j}/2\pi$ & 69.5~GHz & NEMS spectrum\\ \hline
Josephson energy ratio (small/large) & $r$ & 0.16 & NEMS spectrum\\ \hline
Josephson energy (Transmon) & $E_\mathrm{jT}/2\pi$ & 8~GHz$\times$2 & Designed \\ \hline
Coupling strength (NEMS and its readout) & $g_\mathrm{N-r}/2\pi$ & 98~MHz & Bogoliubov fitting\\ \hline
Coupling strength (Transmon and its readout) & $g_\mathrm{T-r}/2\pi$ & 94~MHz & Bogoliubov fitting\\ \hline
Coupling strength (NEMS and Transmon) & $g_\mathrm{N-T}/2\pi$ & 100~MHz & Swap experiment\\ \hline
NEMS frequency (working point) & $\omega_\mathrm{a}/2\pi$ & 5.6~GHz & Rabi and Ramsey\\ \hline
NEMS readout frequency & $\omega_\mathrm{b}/2\pi$ & 7.1~GHz & Direct RF measurement\\ \hline
Transmon frequency (working point) & $\omega_\mathrm{T}/2\pi$ & 3.8~GHz & Rabi and Ramsey\\ \hline
Transmon readout frequency & $\omega_\mathrm{Tr}/2\pi$ & 6.8~GHz & Direct RF measurement\\ \hline
NEMS Kerr (working point) & $K/2\pi$ & 6.9~MHz & Sweep $\ket{1} \leftrightarrow \ket{2}$ transition\\ \hline
NEMS single-photon decay time & $T_\mathrm{1N}$ & $6.0~\mathrm{\mu s}$ & Standard coherence measurement\\ \hline
NEMS transverse relaxation time & $T_\mathrm{2N}$ & $3.0~\mathrm{\mu s}$ & Standard Ramsey measurement\\ \hline
NEMS transverse relaxation time (echo) & $T_\mathrm{2EN}$ & $6.0~\mathrm{\mu s}$ & Standard Ramsey echo measurement\\ \hline
Transmon single-photon decay time & $T_\mathrm{1T}$ & $17~\mathrm{\mu s}$ & Standard coherence measurement\\ \hline
Transmon transverse relaxation time & $T_\mathrm{2T}$ & $1.5~\mathrm{\mu s}$ & Standard Ramsey measurement\\ \hline
Transmon transverse relaxation time (echo) & $T_\mathrm{2ET}$ & $5.4~\mathrm{\mu s}$ & Standard Ramsey echo measurement\\ \hline
NEMS readout resonator linewidth & $\kappa_\mathrm{b}/2\pi$ & 150~kHz & Direct RF measurement\\ \hline
Transmon readout resonator linewidth  & $\kappa_\mathrm{r}/2\pi$ & 300~kHz & Direct RF measurement\\ \hline
Cross-Kerr (NEMS and its readout) & $\chi_\mathrm{N-r}/2\pi$ & 60~kHz & Fitting RF curve, swap experiment\\ \hline
Cross-Kerr (Transmon and its readout) & $\chi_\mathrm{T-r}/2\pi$ & 300~kHz & Fitting the readout curve\\ \hline
Cross-Kerr (NEMS and Transmon) & $\chi_\mathrm{N-T}/2\pi$ & 1.2~MHz & ZZ Ramsey, revival\\ \hline
Frequency-conversion interaction strength & $g_\mathrm{cr}/2\pi$ & 0.23~MHz & Swap experiment\\ \hline \hline
\end{tabular}
\caption{\textbf{Main Parameters.} A summary for the main system parameters. 
More details and additional parameters are discussed in the text.
}
	\label{tab:allparameters}

\end{center}
\end{table}

\section{Calibration of $\Delta_\mathrm{PI,KC}$ and $|\alpha|^2$ using dispersive readout}\label{ZSandalpha}

\subsection{$T_\mathrm{1cat}$ experiment}\label{T1cat}

\begin{figure}[htbp]
	\centering
	\includegraphics{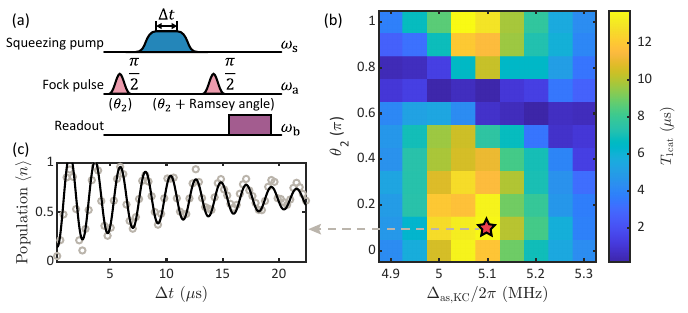}%
	\caption{\textbf{$\textbf{T}_{\textbf{1cat}}$ experiment.} This experiment calibrates the PIFS and the phases of different RF drives ($\omega_\mathrm{a}$ line and $\omega_\mathrm{s}$ line). 
    (a) Pulse sequence.
    (1) Apply a $\pi/2$ pulse with a varying angle $\theta_{2}$ to prepare a superposition state of Fock states on the equator of the Bloch sphere based on \{$\ket{0}$, $\ket{1}$\}, (2) map the state to a Kerr-cat state on the equator of the Kerr-cat Bloch sphere, (3) wait for a variable time $\Delta t$, (4) map back to the Fock state basis, (5) apply another $\pi/2$ pulse with a rotated Ramsey angle, and (6) perform a dispersive readout at the end. The mapping process employs the dynamic compensation method. 
    (b) Experimental results for $T_\mathrm{1cat}$ as a function of $\Delta_\mathrm{as,KC}$ and $\theta_{2}$. The asterisk marks the highest $T_\mathrm{1cat}$. (c) Data and fit for experimental results with different waiting times at the point marked by the asterisk in (b).
 }
	\label{fig:catt1}  
\end{figure}

To calibrate the PIFS, we propose a relatively accurate and straightforward method. Instead of minimizing the Z-rotation rate through tuning the squeezing drive frequency~\cite{KCGrimm2020Nature}, we focus on maximizing the lifetime $T_\mathrm{1cat}$ of a properly chosen $\ket{\alpha}$ state.
The sequence for measuring $T_\mathrm{1cat}$ is shown in Fig.~\ref{fig:catt1}(a). Firstly, the Fock state $\ket{+}=(\ket{0}+\ket{1})/\sqrt{2}$ is prepared via a $\pi/2$ pulse with an angle $\theta_{2}$ in the XY plane and then mapped to $\ket{\alpha}$ with the dynamic compensation method, with the detuning $\Delta_\mathrm{as,KC}$ at the end of the ramping-up time $t_\mathrm{up}$. After waiting for a variable duration, the state is mapped back to the Fock space, followed by the application of another rotated $\pi/2$ pulse and a final average population measurement. By sweeping both $\theta_{2}$ and $\Delta_\mathrm{as,KC}$, the lifetime of $\ket{\alpha}$ $T_\mathrm{1cat}$ is determined from the measured data. 


The longest lifetime, marked by an asterisk in Fig.~\ref{fig:catt1}(b), refers to the optimal angle $\theta_{2}$ for preparing a Kerr-cat state along the X-axis and the proper detuning $\Delta_\mathrm{as,KC}=\Delta_\mathrm{PI,KC}$ to compensate the PIFS. 
The compensation detuning $\Delta_\mathrm{as,KC}/2\pi$ is determined to be 5.1~MHz under this drive strength.
The measured data for the longest lifetime are presented in Fig.~\ref{fig:catt1}(c), showing $T_\mathrm{1cat}=13~\mu$s, which is much longer than the NEMS lifetime $T_\mathrm{1N}=6.0\,\mu$s.  Here, the states $\ket{\pm\alpha}$ cannot be distinguished by measuring the lifetime, leading to a period of $\pi$ for $\theta_2$. Additionally, the angle $\theta_2$ comes from the physical delay of the microwave lines and may vary for different PIFS. 

\subsection{Time Rabi experiment along the X-axis}\label{XRabi}

\begin{figure}[htbp]
	\centering
	\includegraphics{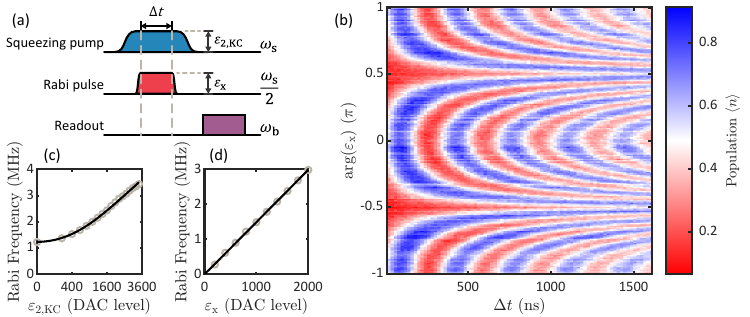}
	\caption{\textbf{X rotation experiment.} This experiment calibrates the cat size and the phases of different RF drives ($\omega_\mathrm{s}/2$ line and $\omega_\mathrm{s}$ line). 
    (a) Pulse sequence.
    (b) Time Rabi results.
    (c) Dependence of the Rabi frequency $\Omega_\mathrm{X}$ on $\varepsilon_\mathrm{2,KC}$. (d) Dependence of the Rabi frequency $\Omega_\mathrm{X}$ on $\epsilon_\mathrm{x}$. 
    In (b) and (c), the strengths of the Rabi pulse are the same with $\abs{\epsilon_\mathrm{x}~(\mathrm{DAC level})} = 2000$ and $\abs{\epsilon_\mathrm{x}}/2\pi = 0.62$ MHz.
    In (b) and (d), the strengths of the two-photon flux pump are the same as that at the asterisk point in Fig.~\ref{fig:catt1}(b), with $\varepsilon_\mathrm{2,KC}~(\mathrm{DAC level}) =2400$ and $\varepsilon_\mathrm{2,KC}/2\pi=9.7$ MHz.
    In (c) and (d), the phases of the two control lines are aligned such that $\arg(\epsilon_\mathrm{x})=0$. 
    }
	\label{fig:catxrabi}  
\end{figure}

After calibrating $\Delta_\mathrm{PI,KC}$, we can perform Rabi experiments on the $\ket{C_\alpha^+}$ state to determine $|\alpha|^2$ for a given flux pump strength~\cite{KCGrimm2020Nature}.
For a Kerr-cat qubit, pulses with a frequency of $\omega_\mathrm{s}/2$ lead to a rotation around the X-axis~\cite{KCPuri2020SciAdvBPCX}. The corresponding sequences are shown in Fig.~\ref{fig:catxrabi}(a). 
Note that, since we have already calibrated the PIFS, we initialize the Kerr-cat state with the dynamic compensation and then implement the Rabi drive using the charge pump line. 
The Rabi oscillation frequency is given by: 
\begin{equation}\label{eq:omegax}
    \Omega_\mathrm{X} = \frac{\Re(4\epsilon_\mathrm{x}\alpha)}{\sqrt{1-\exp(-4|\alpha|^2)}},
\end{equation} 
where $\epsilon_\mathrm{x}$ is the strength of the Rabi drive.


The dependence of the Rabi frequency on the three variables $\arg(\epsilon_\mathrm{x})$, $\varepsilon_\mathrm{2,KC}~\mathrm{(DAC~ level)}$, and $\epsilon_\mathrm{x}~\mathrm{(DAC~ level)}$ are shown in Figs.~\ref{fig:catxrabi}(b), \ref{fig:catxrabi}(c), and \ref{fig:catxrabi}(d), respectively.
In Fig.~\ref{fig:catxrabi}(b), we have the same flux pump strength as that in Fig.~\ref{fig:catt1}.
In Fig.~\ref{fig:catxrabi}(c), the black line is a fit to Eq.~(\ref{eq:omegax}), which gives $|\alpha|^2$ and $\varepsilon_\mathrm{2,KC}$ for each point.
In particular, we get $|\alpha|^2 = 1.41$ and $\varepsilon_\mathrm{2,KC}/2\pi=9.7$~MHz for Fig.~\ref{fig:catt1} and Fig.~\ref{fig:catxrabi}(b, d).
In Fig.~\ref{fig:catxrabi}(d), by fitting the data with a model of rotating around the X-axis and Z-axis at the same time, it gives $\Omega_Z \approx 0$, confirming the good compensation of the PIFS during $\Delta t$. 

\section{Operations in the Kerr-cat space}\label{OperationsKerrCat}


\subsection{Cat Readout}\label{CatReadout}
The cat readout employs a differential frequency signal to swap the state in the NEMS to the readout resonator~\cite{KCGrimm2020Nature,KCAndy2024arxivDetector}. 
During this process, the two-photon squeezing pump remains on, ensuring that the state of the cat qubit is preserved, while the readout resonator is displaced to a coherent state for readout.
 
\begin{figure}[htbp]
	\centering
	\includegraphics{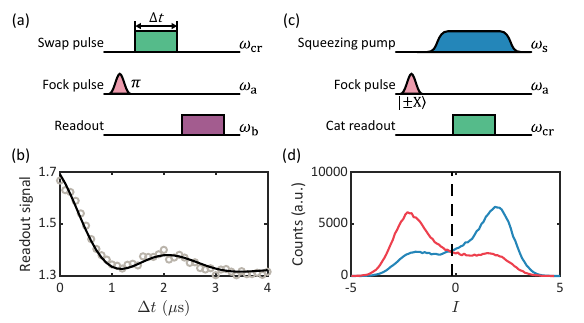}
	\caption{\textbf{Cat readout.} (a) Sequence for measuring $g_\mathrm{cr}$.
    (b) Experimental result and fitting of the swap process. The fitting equation follows that described in Ref.~\cite{KCGrimm2020Nature}. (c) Sequence for the cat readout. (d) Histogram of the cat readout, with each line representing the results of 200,000 experiments.}
	\label{fig:catreadout}  
\end{figure}

 As shown in Figs.~\ref{fig:catreadout}(a-b), the coupling strength is measured to be $g_\mathrm{cr}/2\pi=0.23$~MHz. We choose a readout time of $6~\mathrm{\mu}$s. Figures~\ref{fig:catreadout}(c-d) show the single-shot histogram of the cat readout with a fidelity about $70\%$, which includes both state preparation error and readout error. The readout speed and fidelity are limited by the high quality factor of the readout resonator.
 

For an accurate readout, we first sweep the angle for the initial $\pi/2$ pulse on the Fock state to optimize the readout signal. Proper demodulation allows projecting the signal along the $I$-axis in the phase-space coordinate. The maximum signal of $I$ corresponds to $\ket{+\alpha}$, while the minimum signal corresponds to $\ket{-\alpha}$. 

\subsection{X/2 gate}

X($\pi/2$) gate (or X/2 gate) can be directly derived from the Rabi experiment along the X-axis. 
By setting the Rabi pulse to a Gaussian shape and fixing its duration, the calibration involves adjusting the pulse power to achieve the desired rotation.

\begin{figure}[htbp]
	\centering
	\includegraphics{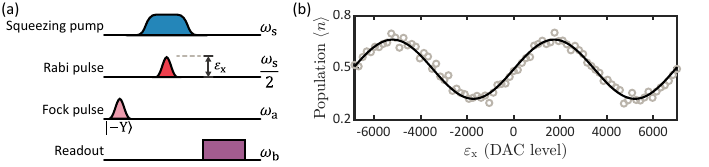}
	\caption{\textbf{X/2 gate.} (a) Sequence to calibrate the X/2 gate. (b) Experimental results and fit using a sinusoidal function.}
	\label{fig:xgate}  
\end{figure}

As shown in Fig.~\ref{fig:catxrabi}(b), the Rabi oscillations as a function of $\theta_\mathrm{x}$ have a period of $\pi$. This makes it impossible to distinguish the rotating direction if the sequence in Fig. S7(a) is adopted. Therefore, we apply a $\pi/2$ pulse between $\ket{0}$ and $\ket{1}$ at the beginning of the sequence, as shown in Fig.~\ref{fig:xgate}(a).
The angle of this pulse is determined through the above cat readout calibration experiment. As a result, we prepare a $\ket{+\alpha}$ state, and then sweep the power of the Rabi pulse. A sinusoidal function is used to fit the curve. 
If the rotation direction is incorrect, the fit will yield a negative sinusoidal function.

\subsection{Z/2 gate}

\begin{figure}[htbp]
	\centering
	\includegraphics{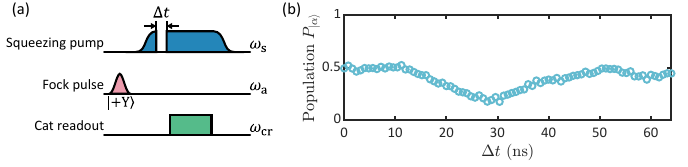}
	\caption{\textbf{Z/2 gate.} (a) Sequence to calibrate the Z/2 gate. (b) Experimental results. The time corresponding to the minimum population of $\ket{\alpha}$ is considered the optimal time for the Z/2 gate.}
	\label{fig:zgate}  
\end{figure}

Z($\pi/2$) gate (or Z/2 gate) can be realized by turning off the two-photon squeezing pump for a certain time, during which the state will go out of the Kerr-cat computational space. 
Since the readout signal is always bounded by those for $\ket{\pm\alpha}$, regardless of whether the state is inside or outside the Kerr-cat computational space \cite{KCAndy2024arxivDetector}, we can determine the appropriate gate time from the cat readout values. As shown in the experimental sequence in Fig.~\ref{fig:zgate}(a), we first prepare a $\ket{\mathrm{+Y}}$ $(\ket{C_\alpha^{-\mathrm{i}}})$ with a $\pi/2$ gate in the Fock space. A Z/2 gate will rotate it to $\ket{-\alpha}$, resulting in a minimum signal as in Fig.~\ref{fig:zgate}(b). The corresponding time refers to the optimal time for the Z/2 gate.

\section{Characterization of the Mapping Process}\label{Characterization of Map Process}
\subsection{Quantum Process Tomography}
\label{Process Tomography}
We carry out quantum process tomography (QPT) to characterize the initialization operation. 
In this work, we choose R matrix (Pauli-transfer-matrix) representation for the QPT~\cite{FidelityRChow2012PRL,KCGrimm2020Nature}. A standard procedure of QPT used in Ref.~\cite{NielsenChuang} is based on the first two Fock states $\ket{0}$ and $\ket{1}$. Fidelity, the final result of QPT, is related to projections on each axis of the Bloch sphere. Ref.~\cite{KCGrimm2020Nature} takes this method to characterize the fidelity of the mapping operation.

Obviously, this procedure requires to prepare a set of Fock states and measure the final states also in the Fock space. This means that we have to add a symmetric ramping-down process, otherwise the final quantum states are in the Kerr-cat space. However, this whole mapping operation behaves differently from the one with only the initialization part. 
For instance, in the main text we discuss that the static compensation method may adiabatically map the vacuum state to a leakage state. However, an exactly symmetric ramping-down process will map the leakage state back to its original state. In this situation, the whole mapping operation behaves like an ideal identity channel, which has been discussed in Fig.~3 in the main text. Nonetheless, despite the effects of leakage, this procedure can still illustrate the influence of decoherence.

A more effective approach is to define effective qubit density matrices both in the Fock space and the Kerr-cat space~\cite{KCIyama2024NCWigner}. A similar procedure can then be employed to characterize solely the initialization process. Using the X/2 and Z/2 gates, we can measure the projections along each axis of the Kerr-cat Bloch sphere. The following are the specific details for the Pauli-transfer-matrix method and fidelities in the two spaces.

\subsubsection{Pauli-transfer-matrix Method}\label{PTM}

A density matrix of a qubit can be written as
\begin{equation}\label{eq:PIXYZ}
    \rho = \frac{1}{2}(P_II+P_XX+P_YY+P_ZZ),
\end{equation}
where $P_S=\mathrm{Tr}(\rho S)$ for $S=I,\,X,\,Y,\,Z$, and $I,\,X,\,Y,\,Z$ are the identity and Pauli matrices. Generally, $P_I=1$ and a deviation from unity represents the leakage.

A ``p vector" can be defined in the Pauli basis as:
\begin{equation}\label{eq:pvector}
    \overrightarrow{p} = 
        \left[\begin{array}{c}
        P_I \\
      	P_X \\
       	P_Y \\ 
       	P_Z 
        \end{array}\right].
\end{equation}
A process can be treated as a mapping from an input vector to an output one, meaning a matrix $R$ connects the initial and final density matrices:
\begin{equation}\label{eq:pRp}
    \overrightarrow{p}_\mathrm{out} = R\overrightarrow{p}_\mathrm{in},
\end{equation}
where $\overrightarrow{p}_\mathrm{in}$ and $\overrightarrow{p}_\mathrm{out}$ are the p vector form of the initial and final density matrices. $R$ is the Pauli-transfer-matrix (or ``R matrix") representation of the quantum process. Notice that the elements of $R$ must be real. Then, fidelity is defined as~\cite{FidelityRChow2012PRL}
\begin{equation}\label{eq:fidelityR}
    F_R = \frac{Tr(R_\mathrm{ideal}^{\dagger}R_\mathrm{exp}+d)}{d^2+d},
\end{equation}
where $R_\mathrm{ideal}$ and $R_\mathrm{exp}$ are the ideal and experimental R matrix, respectively, and $d=2$ is the dimension of the density matrix (or effective density matrix in the Kerr-cat space).

\subsubsection{Fidelities of the mapping operation in the Fock space}\label{FockFidelities}

\begin{figure}[htbp]
	\centering
	\includegraphics{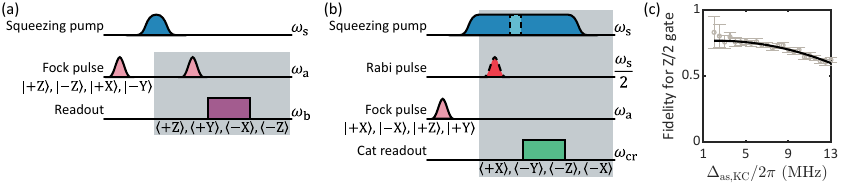}
	\caption{(a) Sequence for QPT in the Fock space. 
	(b) Sequence for QPT in the Kerr-cat space. 
	(c) Fidelity of the Z/2 gate for different PIFS in experiment. The black curve represents a polynomial fit of the experimental results, which we utilize in our QuTiP simulation (``Sim., SPAM" in Fig.~4 of the main text).
	}
	\label{fig:Tomo}  
\end{figure}

The sequence for QPT in the Fock space is shown in Fig.~\ref{fig:Tomo}(a). First, we prepare four initial states $\ket{+Z}=\ket{0},~\ket{-Z}=\ket{1},~\ket{+X}=\ket{+} = (\ket{0}+\ket{1})/\sqrt{2}$, and $\ket{-Y}=\ket{-\mathrm{i} } = (\ket{0}-\mathrm{i} \ket{1})/\sqrt{2}$. Then, we measure the final states along $+Z,\,+Y,\,-X,\,-Z$ axes to obtain the R matrix. Notably, to mitigate the impact of the leakage out of the encoding space, we do not perform maximum likelihood estimation (MLE). We prepare $\ket{1}$, $\ket{+}$, and $\ket{-\mathrm{i}}$ using a $\pi$ pulse or $\pi/2$ pulse with different angles (called ``initial pulse"). Following that, we execute the ramping-up and ramping-down mapping operations. For projection measurements, we apply a $\pi$ or $\pi/2$ pulse (called ``tomography pulse") to rotate the target axis to +Z axis and measure the population of $\ket{0}$ $P_{\ket{0}}$.


{In the experiment, single-shot readout for each Fock state is challenging, because the dispersive shift is small and the readout signals for Fock states $\ket{0},\,\ket{1},\,\ket{2},...$ are nearly linear with Fock number. Therefore, we employ approximate methods to process the measured data.} Here, we assume the population of the leakage state is much smaller than 1. With this assumption, for each initial state in experiment we calculate $[P_n(+Z)+P_n(-Z)-1]/2$ as a leakage offset of the readout result $P_n$. Subtracting this offset, we approximately get $P_{\ket{1}}$ and further $P_{\ket{0}}\approx1-P_{\ket{1}}$. Using these $P_{\ket{0}}$ results, we can calculate the ``p vector".
This data processing ensures $P_{\ket{0}}\leq1$ while retaining the impact of the leakage. Future work could focus on improving the readout fidelity.

We note that the simulation results with QuTiP~\cite{qutip1,qutip2} presented in Fig.3 of the main text also follow the above procedure. We first calculate the offset induced by leakage, then determine the approximate $P_{\ket{0}}$, and finally obtain the R matrix and fidelity.

\subsubsection{Fidelities of the initialization in the Kerr-cat space}\label{CatFidelities}

Performing QPT in the Kerr-cat space is similar to that in the Fock space [Fig.~\ref{fig:Tomo}(b)]. Considering the bias-preserving axis is X, we change the four initial states to $\ket{+X},\,\ket{-X},\,\ket{+Z}$, and $\ket{+Y}$, which are $\ket{\alpha},\,\ket{-\alpha},\,\ket{C_{\alpha}^+}$, and $\ket{C_{\alpha}^{-\mathrm{i}}}$ in the Kerr-cat space. These states correspond to $\ket{+} = (\ket{0}+\ket{1})/\sqrt{2},~\ket{-} = (\ket{0}-\ket{1})/\sqrt{2},~\ket{0}$, and $\ket{\mathrm{i} } = (\ket{0}+\mathrm{i} \ket{1})/\sqrt{2}$ in the initial Fock space. In this situation, the measurement axis should also be changed correspondingly. 
After the Kerr-cat initialization, four projections on +X, -X, +Z, +Y axes are measured as follows.  
+X: perform no operations and measure population of $\ket{\alpha}$;
-X: perform no operations and measure population of $\ket{-\alpha}$;
+Z: perform X/2 and Z/2 gates and measure population of $\ket{-\alpha}$;
+Y: perform Z/2 gate and measure population of $\ket{\alpha}$.

We take the average value of the cat readout signals to calculate the population. Note that higher-level leakages will also contribute to the readout signal of $\ket{\alpha}$ and $\ket{-\alpha}$. Therefore, it is essential to know the influence of the higher-level states for our readout signal, as they cannot be distinguished. These higher-level states will go through two possible physical processes to affect the final result. On one hand, due to the relatively long measurement time, the higher-level states may decay back to the Kerr-cat space and be detected. However, for our cat size, the higher-level states are non-degenerate, making them more likely to cause information transfer from one potential well to another~\cite{KDCGautier2022PRXQ}. This mixing of states ultimately results in an intermediate readout signal after decaying back to the encoded space. 

On the other hand, even in the absence of decay, the elevated states beyond the double potential wells will yield an intermediate measurement outcome. Similar results of performing cat readout to higher-level states have been observed in another recent work \cite{KCAndy2024arxivDetector}. Both of these two effects make the measured fidelity converge to 0.5. Therefore, compared to QPT in the Fock space, which is difficult to distinguish whether having leakage states or not, this approach may serve as a more effective indicator for comparing different methods of Kerr-cat initialization.

To get the fidelity curves for different methods, the procedure is as follows:

1. Calibrate parameters for JPA, NEMS frequency, $\pi$ pulse strength, etc.

2. Perform cat readout experiment with different angles of the preparation $\pi/2$ pulse to calibrate the relative phase between the flux pump line and the charge drive line.

3. Prepare $\ket{\alpha}$ and calibrate the PIFS for a specific squeezing amplitude using $T_\mathrm{1cat}$ experiment.

4. Perform cat readout experiment again to calibrate the demodulation angle.

5. Calibrate the angle and power of the X rotation in the Kerr-cat space.

6. Calibrate the gate time for the Z/2 gate.

7. Measure the fidelity as mentioned above.

This procedure covers almost all of the experiments that are described previously in this Supplementary Materials. The results are mainly limited by the fidelity of the Z/2 gate. 
In Fig.~\ref{fig:zgate}, we see a Z/2 gate cannot reach the lowest boundary of the population, where $P_{\ket{\alpha}}=0$. When both the pump strength and the PIFS increase, the situation becomes even worse. Experimental fidelity of the Z/2 gate is shown in Fig.~\ref{fig:Tomo}(c). The fidelity is defined as the population of the target state, for example $P_{\ket{-\alpha}}$ for Fig.~\ref{fig:zgate}. We employ a polynomial fit [Fig.~\ref{fig:Tomo}(c), black line] to generate a fidelity curve through simulation in the presence of a noisy Z/2 gate (solid lines in Fig.~4 of the main text).

There are alternative ways to perform the Z/2 gate, such as mapping back to the Fock space and numerically adding a virtual $\pi/2$ phase. However, considering this work mainly focuses on the mapping process, we do not take this method. On the contrary, this is exactly one of the applications of this work: we use the dynamic compensation method for the mapping process, and then realize an arbitrary Z rotation for a Kerr-cat qubit.

\subsection{Wigner Function}
\label{Wigner}

While process fidelity provides a comprehensive indicator for the transfer of the entire Bloch sphere, what we concern about mostly in this work is the mapping of $\ket{0}\rightarrow\ket{C_{\alpha}^+}$, because this mapping process is not subject to gate fidelity in the Fock space. Moreover, a measurement after the mapping with post-selection collapses $\ket{C_{\alpha}^+}$ to $\ket{\alpha}$ or $\ket{-\alpha}$, which is a convenient way to prepare bias-preserving coherent states~\cite{KCGrimm2020Nature,KCFrattini2022SpectralKissing,KCAndy2024arxivDetector}.



\begin{figure}[htbp]
	\centering
	\includegraphics{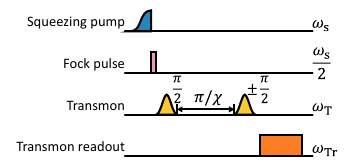}
	\caption{\textbf{Sequence for measuring Wigner Functions.} 
    It is worth noting that the length of the displacement pulse is 4.8~ns, which is much smaller than $1/(20K)$, hence satisfying the Kerr evolution correctable condition mentioned in Ref.~\cite{KCIyama2024NCWigner}.
	In the figure, ``$\pm\frac{\pi}{2}"$ means that two steps are applied, with the second $\pi/2$ pulse either in the same direction or the reverse direction of the first one, and we take their average as the final result for the parity. 
	}
	\label{fig:wigner}  
\end{figure}

The Wigner function offers a more intuitive way to characterize the $\ket{C_{\alpha}^+}$ state that we prepare. Moreover, Wigner function can directly reflect the leakage issue of the initialization process, especially when PIFS is larger than Kerr. We measure the Wigner function of the NEMS, following the methodology in Ref.~\cite{KCIyama2024NCWigner} . The measurement sequence is shown in Fig.~\ref{fig:wigner}. The Wigner function is defined as
\begin{equation}
    W(\alpha) = \frac{2}{\pi}\mathrm{Tr}(\hat{D}^\dagger(\alpha)\rho\hat{D}(\alpha)\hat{P}),
\end{equation}
where $\hat{D}(\alpha)=\exp(\alpha\hat{a}^\dagger-\alpha^*\hat{a})$ is the displacement operator, $\hat{P}=\exp(\mathrm{i}\pi\hat{a}^\dagger\hat{a})$ is the photon number parity operator, and $\rho$ is the density matrix.

The traditional way to measure the Wigner function is to apply a resonant drive as a displacement operation on a harmonic oscillator (typically a cavity) that is dispersively coupled to a two-level system (typically a transmon) with a coupling strength $\chi$. Afterward, the parity of the harmonic oscillator is measured through the two-level system~\cite{ParitySun2014Nature}.
The principle to measure parity is that the transmon accumulates different phases depending on the state of the cavity. 
If the wait time is set to $\pi/\chi$, the transmon will transition to either a ground or excited state depending on the even or odd parity state of the cavity.

However, the situation is difficult when the harmonic oscillator has a small but finite Kerr nonlinearity. The resonant drive is no longer a displacement operation. Nevertheless, when the time duration is sufficiently short and the power is sufficiently high, we can approximate the drive as a small rotation induced by Kerr, followed by a displacement. Consequently, what we measure is a distorted Wigner function, which can be corrected back with a unitary operator $\hat{U}_\mathrm{cor}$ once we get the density matrix after the Kerr rotation~\cite{KCIyama2024NCWigner}. Here, 
\begin{equation}\label{eq:Ucor}
    \hat{U}_\mathrm{cor} = \exp(-\frac{\mathrm{i}}{\hbar}\frac{Kt_\mathrm{cor}}{2}\hat{a}^\dagger\hat{a}^\dagger\hat{a}\hat{a}),
\end{equation}
where $t_\mathrm{cor}$ is the rotation time which can be chosen properly with the help of simulation. The final result will be
\begin{equation}\label{rhowigner}
    \rho = \hat{U}_\mathrm{cor}\rho_0\hat{U}_\mathrm{cor}^\dagger,
\end{equation}
where $\rho_0$ is the density matrix with the direct measurement of the Wigner function. There are different ways to transfer Wigner functions to density matrices, such as conditional generative adversarial network~\cite{CGANAhmed2021PRL,CGANAhmed2021PRR} (in Ref.~\cite{KCIyama2024NCWigner}) or MLE (in this work).

In Fig.~\ref{fig:wigner}, we perform $\pm\frac{\pi}{2}$ gates for the second transmon pulse. To minimize system errors in the tomography, we collect the data from both experiments and calculate the average to get the final parity results.

To sum up, the procedure to measure a Wigner function is as follows:

1. Calibrate the PIFS for a specific squeezing amplitude.

2. Initialize a Kerr-cat cat state with different compensation methods.

3. Apply a short and strong resonant charge drive on the NEMS and carry out parity measurements with the transmon.

4. Get the Wigner function and perform MLE to get the density matrix.

5. Correct the density matrix with proper unitary operation.

6. Obtain Wigner function based on the corrected density matrix.




%